\documentclass[12pt]{article}
\pdfoutput=1
\usepackage{amsmath, amssymb}
\usepackage{graphicx}
\usepackage{graphicx}
\usepackage{color}

\newcommand{\beq}{\begin{equation}}
\newcommand{\eeq}{\end{equation}}
\newcommand{\beqa}{\begin{eqnarray}}
\newcommand{\eeqa}{\end{eqnarray}}
\newcommand{\beqar}{\begin{eqnarray*}}
\newcommand{\eeqar}{\end{eqnarray*}}

\newcommand{\bea}{\begin{eqnarray}}
\newcommand{\eea}{\end{eqnarray}}

\newcommand{\ie}{{\it i.e.,}\ }
\newcommand{\reef}[1]{(\ref{#1})}
\newcommand\cO{{\cal O}}

\begin{document}

\vspace*{1cm}

\begin{center}
{\bf \Large  Charged Rotating AdS Black Holes with \\
\vspace*{.5cm}
 Chern-Simons Coupling}

\vspace*{.5 cm}

{{Mozhgan Mir}$^{\ast,\dagger,}$\footnote{mmir@ipm.ir}} and {{Robert B. Mann}$^{\ast,}$\footnote{rbmann@uwaterloo.ca}}
\\
\vspace*{1cm}
\textit{
$ {}^{\ast}$ Department of Physics and Astronomy, University of Waterloo,\\
Waterloo, Ontario, Canada, N2L 3G1\\
\vspace*{.5cm}
$ {}^{\dagger}$School of Physics, Institute for Research in Fundamental Sciences (IPM),\\
P.O.Box 19395-5531, Tehran, Iran
}
\\
\vspace{2cm}

\end{center}

\begin{abstract}
\baselineskip=18pt
We obtain a perturbative solution for  rotating charged black holes in 5-dimensional Einstein-Maxwell-Chern-Simons theory with a negative cosmological constant. We start from a small undeformed Kerr-AdS solution and use the electric charge as a perturbative parameter to build up black holes with equal-magnitude angular momenta up to forth order. These black hole solutions are described by three parameters, the charge, horizon radius and horizon angular velocity. We determine the physical quantities of these black holes and study their dependence on the parameters of black holes and arbitrary Chern-Simons coefficient. In particular, for values of CS coupling constant beyond its supergravity amount, due to a rotational instability, counterrotating black holes arise. Also the rotating solutions appear to have vanishing angular momenta and do not manifest uniquely by their global charges.

\end{abstract}

\vfill
\setcounter{page}{0}
\setcounter{footnote}{0}
\newpage

\section{Introduction } \label{intro}
Higher dimensional black holes have been subject of interest in recent years, particularly black hole solutions in asymptotically AdS spacetime due to their applications via the AdS/CFT conjecture to build the corresponding elements of the field theory \cite{Witten:1998}  (for an application see \cite{Mir:2013pca} and references therein).  The original higher-dimensional
version of the Schwarzschild solution \cite{Tangherlini:1963} was later extended to include rotation \cite{Myers:1986}, and then further generalized to included a cosmological constant \cite{Pope:2004}.
In $D$-dimensions, these solutions have $N=\left[\frac{D-1}{2}\right]$ independent  spatial planes of rotation and N independent angular momenta.

One particular theory of interest is  Einstein-Maxwell-Chern-Simons theory (EMCS) with a negative cosmological constant in five-dimensions. For a special value of the Chern-Simons (CS) coupling $\lambda=\lambda_{SG}=1/(2 \sqrt{3})$, this theory is equivalent to 5-dimensional minimal gauged supergravtiy \cite{Pope:2005}. However for arbitrary values of  $\lambda$ the exact rotating charged black hole solutions in five dimensions are not known.  In the extremal case they have static horizons $\Omega_H=0$  but nonzero angular momentum $J$ \cite{Myers:1997}.

In the absence of analytic solutions, one can follow either perturbative
or numerical approaches.
The closed form of rotating AdS EM black holes have been analyzed numerically \cite{Navarro:2007B,Brihaye:2009} and extended to include a Gauss-Bonnet term \cite{Cai:2008}.
Numerical solutions for asymptotically flat  EMCS black holes, in the case of spherical topology and two equal angular momenta, have also been found along with their extremal black hole counterparts
\cite{Kunz:2005,Kunz:2006B}.

Such charged rotating black holes have a number of interesting features.
  For $\lambda_{SG}<\lambda<2\lambda_{SG}$  the mass can decrease by increasing the angular momentum \cite{Kunz:2005}. The extremal static black hole with zero Hawking temperature cannot emit energy and angular momentum via Hawking radiation;  the angular momentum  accumulates behind the horizon and causes rotational instability and a deformation of the black hole \cite{Myers:1999,Herdeiro:2000}.   In terms of the first law of thermodynamics, for extremal black holes with vanishing temperature and fixed charge ($dQ=0$), this instability causes the horizon to rotate in the opposite direction of the angular momentum, a phenomenon called  counterrotation
\cite{Kleihaus:2004}; since $dM=\Omega_H dJ$,   an increase in $J$ causes a decrease in $M$ \cite{Myers:1999}.
For $\lambda>2\lambda_{SG}$, in addition to the static Reissner$-$Nordstrom solution, there are two extremal stationary solutions with vanishing global angular momenta
\cite{Kunz:2005}.  The analysis of these cases is carried out by using a modified near-horizon formalism through inclusion of the CS term where the near horizon solutions are asymptotically flat  \cite{Blazquez-Salcedo:2013muz,Kunz:2015}
These black holes are not uniquely characterized by their global charges.

From a perturbative perspective,  the gyromagnetic ratio for charged Kerr-AdS  black holes was computed to leading order in the charge \cite{Aliev:2007}, and slowly rotating charged asymptotically flat black hole solutions were obtained by perturbing in terms of the angular momentum  \cite{Aliev:2006,Aliev:2006D}. Asymptotically flat charged rotating black holes in five dimensional Einstein-Maxwell and EMCS theories to higher order in the charge $Q$ for  equal-magnitude of angular momenta have also been obtained  \cite{Navarro:2007,Kunz:2010a04}.

Here we construct  general  rotating AdS black holes in EMCS theory in 5 dimensions for arbitrary values of the CS coupling constant $\lambda$. We focus on black hole solutions with equal-magnitude angular momenta,  employing the black hole charge as an expansion parameter. Related results using  this approach for  asymptotically flat extremal solutions have been obtained \cite{Kunz:2010a04}.  However our solutions form a three parameter family comprised of charge, horizon radius and horizon angular velocity, whose radii are small compared to the $AdS_5$ radius.  The black hole uniqueness  theorem in 4 dimensions states that asymptotically flat non-degenerate electrovac black holes are uniquely characterized by their global charges \cite{Israel:1968,Robinson:1975,Mazur:1982,Heusler:2012}. We observe at  order $Q^3$ in the charge that uniqueness  is not satisfied, commensurate with previous work on EMCS theory \cite{Blazquez-Salcedo:2013muz} . Indeed we find that expanding to this order is sufficient to see many interesting features of these black holes, including various thresholds in the Chern-Simons couplings that introduce new behaviour in the various physical quantities of these black holes,
counter-rotation perhaps being the most prominent example.

 An outline to our paper is as follows. In section \ref{EOM} we introduce the ansatz for the metric and the gauge potential for 5-dimensional black holes with equal-magnitude angular momenta.
In section \ref{matching} we demonstrate the known analytic solution for $\lambda_{SG}$ where this solution is rewritten in the coordinates given in \cite{Dias:2011at}. Section \ref{construction} explains the perturbation method we employ for obtaining rotating charged black holes with arbitrary CS coefficient
$\lambda$.
In section \ref{thermo} we present the expressions for the physical quantities of these rotating charged black holes: their  energy, angular momentum, surface gravity and horizon area. Section \ref{explor} is devoted to analyze how these different physical quantities depend on the black hole parameters. Details of the field equations we employ with our ansatz and our perturbative solutions are given in appendices.
For completeness we present the corresponding perturbative expansions from the exact known
supergravity solution \cite{Pope:2005} in appendix \ref{pertpope}; we find our perturbative result fully consistent with this expansion to the order in which we work.

\section{Basic equations of motion} \label{EOM}

The action for Einstein-Maxwell-Chern-Simons theory in 5 spacetime dimensions is
\beqa
S &=& \frac{1}{16 \pi}\int_{{\cal M}} d^5 x \sqrt{-g}\left[R-2\Lambda-F_{a b}F^{a b}+\frac{4}{3}\lambda \epsilon^{a b c d e}F_{a b}F_{c d} A_e
\right]
\label{action}
\eeqa
where the cosmological constant $\Lambda = -6/l^2$ and the Chern-Simons coupling is $\lambda$.   For $\lambda= \lambda_{SG}$ the action for five-dimensional gauged supergravity is recovered.\\
The field equations are given by
\beqa
G_{a b}-\frac{6}{ l^2}g_{a b}=2{T}_{a b}, \quad \quad
\nabla_{b}F^{b a}+ \lambda \epsilon^{a b c d e}F_{b c}F_{d e }=0
\label{Eineq}
\eeqa
where
\beqa
{T}_{a b}&=&F_{a d}{F_{b}} ^{d}-\frac{1}{4}g_{a b} F_{c d}F^{c d}
\label{Tab}
\eeqa
is the stress energy tensor for the gauge field.

We seek perturbative rotating black hole solutions to these equations by deploying the ansatz
\beqa
ds^2&=&-f(r) g(r) dt^2+\frac{dr^2}{f(r)}+r^2\left[h(r)\left(d\psi+\frac{\cos \theta}{2}d\phi-\Omega(r) dt\right)^2+\frac{1}{4}(d\theta^2+\sin^2\theta d\phi^2)\right]\nonumber\\
\label{metric}
\eeqa
for the metric and
\beqa
A(r)&=&a_{0}(r)dt+a_{1}(r)\left(\frac{1}{2}\cos(\theta)d\phi+d\psi\right)\label{Aansatz}
\eeqa
for the gauge field.
This ansatz, corresponding to a doubly-rotating black hole whose angular momenta are equal,
ensures that the Einstein-AdS-Chern-Simons equations \eqref{Eineq} reduce to a set of 6 ordinary differential equations for the 6 unknown functions $\{f,g,h,\Omega,a_0,a_1\}$.

In appendix A we present the field equations \eqref{Eineq} that result upon using our ansatz
(\ref{metric},\ref{Aansatz}) for the metric and gauge field.\
We are specifically interested in those solutions to \eqref{Eineq} that asymptote to AdS$_5$ spacetime, $\ie$ solutions whose large-$r$ behaviour is given by
\beqa
f |_{r\rightarrow \infty}&=&\frac{r^2}{l^2}+1+\frac{C_f l^2}{r^2}+{{\cal}O}(r^{-3}), \quad \quad  g |_{r\rightarrow \infty}=1-\frac{C_h l^2}{r^2}+{{\cal}O}(r^{-5}), \nonumber\\
h |_{r\rightarrow \infty}&=& 1+\frac{C_h l^2}{r^2}+{{\cal}O}(r^{-5}), \quad \quad  \Omega |_{r\rightarrow \infty}=\frac{C_{\Omega} l^3}{r^4}+{{\cal}O}(r^{-4}),  \label{metbc}\\
a_0 |_{r\rightarrow \infty}&=&\frac{C_{a_0} l^2}{r^2}+{{\cal}O}(r^{-3}), \quad\quad \quad a_{1} |_{r\rightarrow \infty}=\frac{C_{a_1} l^3}{r^2}+{{\cal}O}(r^{-3}) \nonumber
\eeqa
For the black hole, the inner boundary is at its horizon where the function $f(r)$ must vanish. We take this condition as our definition for the location of the black hole horizon $r_{+}$ (the largest root of $f(r)$). The other functions must be regular at this hypersurface:
\beqa
f\big|_{r \rightarrow r_{+}} = \mathcal{O}(r-r_{+}), &\quad
g\big|_{r \rightarrow r_{+}} = g(r_+)+\mathcal{O}(r-r_+), \quad
h\big|_{r \rightarrow r_{+}} = h(r_+)+\mathcal{O}(r-r_+)  \label{bndry} \\
\Omega\big|_{r \rightarrow r_{+}} = \Omega_H+\mathcal{O}(r-r_+), &\quad
a_0 |_{r\rightarrow r_{+}}=a_0(r_+)+\mathcal{O}(r-r_+), \quad  a_{1} |_{r\rightarrow r_{+}}=
a_1(r_+)+\mathcal{O}(r-r_+)
\nonumber
\eeqa
These conditions fix some of constants of integration. %We will also be interested in solutions that are regular (in a suitable sense) in the interior.  \tcr{\bf [Have these been computed?]}

\section{Exact Solution} \label{matching}

One well-known solution to these equations is the metric  \cite{Pope:2005}
%\tcr{\bf [Has this been fixed so that notation is consistent with original Pope solution? ]}
\beqa
ds^2&=&
-\frac{\Delta_{\bar\theta}[(1+g^2 \bar r^2)\rho^2 d t+2q \nu] d t}{\Xi_a \Xi_b \rho^2}
+\frac{2q\nu\omega}{\rho^2}+\frac{f}{\rho^4}\left(\frac{\Delta_{\bar\theta}d t}{\Xi_a \Xi_b}-\omega\right)^2
+\frac{\rho^2}{\Delta_{\bar r}}d\bar r^2+\frac{\rho^2}{\Delta_{\bar\theta}}d\bar\theta^2\nonumber\\
&&\left.
+\frac{\bar r^2+a^2}{\Xi_a}\sin^2\bar\theta d\bar\phi^2+\frac{\bar r^2+b^2}{\Xi_b}\cos^2\bar\theta d\bar{\psi}^2\right.\label{popemetric}\\
A&=&\frac{\sqrt{3}q}{\rho^2}\left(\frac{\Delta_{\bar\theta} dt}{\Xi_a \Xi_b}-\omega\right)\nonumber
\eeqa
valid for $\lambda= \lambda_{SG}$, where
\beqa
\nu&=&b\sin^2\bar\theta d\bar\phi+a\cos^2\bar\theta d\bar{\psi},\nonumber\\
\omega&=&a\sin^2\bar\theta\frac{d\bar\phi}{\Xi_a}+b\cos^2\bar\theta\frac{d\bar{\psi}}{\Xi_b},\nonumber\\
\Delta_{\bar\theta}&=&1-a^2 g^2 \cos^2 \bar\theta-b^2 g^2 \sin^2 \bar\theta,\nonumber\\
\Delta_{\bar r}&=&\frac{(\bar r^2+a^2)(\bar r^2+b^2)(1+g^2 \bar r^2)+q^2+2abq}{\bar r^2}-2m\nonumber\\
\rho^2&=&\bar r^2+a^2\cos^2\bar\theta+b^2\sin^2\bar\theta,\nonumber\\
\Xi_a&=&1- a^2 g^2,~~\Xi_b=1- b^2 g^2,\nonumber\\
f&=&2m\rho^2-q^2+2abq g^2\rho^2\nonumber
\eeqa
where  $g$ is related to the radius of AdS space $g=1/l$ and $q$ is  related to
the charge associated with the above gauge potential.

We are interested in finding charged doubly-rotating equal angular momenta solutions to the field equations, for which the rotation parameters $b=a=a_S$, and write $\Xi_S = 1-\frac{a_S^2}{l^2}$. To do this we must express the solution  \eqref{popemetric} in terms of the coordinates
given in \eqref{metric}.  Using the coordinate transformation
\beq
r^2 = \frac{\bar{r}^2 + a_S^2}{\Xi_S} \qquad {\bar\phi}= \psi-\phi/2 \qquad {\bar\psi} = \psi+\phi/2\nonumber
\eeq
with ${\bar\theta}=\theta/2$, we obtain
\bea
f(r) &=& \frac{r^2}{l^2}+1-\frac{2m}{\Xi_S^2 r^2}  + \frac{2Q a_S^2}{r^4 \Xi_S}
+\frac{\Xi_S Q^2}{r^4} +\frac{2 a_S^2 m}{\Xi_S^3 r^4}\nonumber\\
h(r)&=&\frac{1}{g(r)} = 1 +\frac{2 Q a_S^2}{r^4\Xi_S} -\frac{Q^2 a_S^2}{r^6}+\frac{2 a_S^2 m}{\Xi_S^3 r^4}
\label{PopeKAdS}\\
a_S\Omega(r)  &=& 1- \frac{r^4 + a_S^2 Q}{r^4 h(r)} = \frac{a^2_S}{r^6 \Xi^3_S h(r)} \left(2 m r^2 +\Xi_S Q r^2(1-\frac{a_S^4}{l^4})-Q^2 \Xi_S^3\right)
\nonumber
\eea
and where the gauge potential is
\beqa
A=\frac{\sqrt{3}Q}{r^2}\big(dt-a_S(d\psi+\frac{\cos \theta}{2}d\phi)\big)\nonumber
\eeqa
with $q=Q\Xi^2_S$.\\

Setting $a_S=0$ we recover the Reissner-Nordstrom-AdS (RN-AdS) solution.   Setting $Q=0$ we recover
 the doubly-rotating Kerr-AdS (KAdS) metric  \cite{Pope:2004}
\beqa
f_0(r)&=&1+\frac{r^2}{l^2}-\left(1-\frac{a_M^2}{l^2}\right)\frac{r_M^2}{r^2}+\frac{a_M^2 r_M^2}{r^4}, \quad  h_0(r)=1+\frac{a_M^2 r_M^2}{r^4} \nonumber\\
\Omega_0(r)&=&\frac{a_M\ r_M^2}{r^4 h_0(r) } , \quad
g_0(r)=\frac{1}{h_0(r)} \label{KAdSS}
\eeqa
whose two angular momenta are equal, with $a_M=a_S|_{Q=0}$ and  $r_M^2 = (2 m/\Xi_S^3)|_{Q=0}$.   %It is straightforward to verify that    the following angular transformations
%\beq
%{\bar\phi}= \psi-\phi/2 \qquad {\bar\psi} = \psi+\phi/2 \qquad {\bar\theta}=\theta/2
%\eeq
%transform the metric  \tcb{\eqref{popemetric} to the standard form of the Kerr-AdS metric}.

Note that we can rewrite $r_M$ and $a_M$ in terms of $r_+$ and
$\Omega_H$
\beqa
r_M^2=\frac{r_{+}^4 \left(l^2+r_{+}^2\right)}{l^2 r_{+}^2-a_M^2 \left(l^2+r_{+}^2\right)}, \quad
a_M =\frac{r_{+}^2 l^2  \Omega_{H}}{l^2+r_{+}^2} \label{aM}
\eeqa
There are three free parameters for charged equally rotating black holes: the mass parameter $m$, the rotation parameter
$a$ and the charge parameter $Q$. We shall rewrite the former two quantities in terms of the horizon velocity $\Omega_H$ and the horizon radius $r_+$.
We can repeat a similar analysis for \eqref{PopeKAdS}, from  the condition that $f(r)$ defined in \eqref{popemetric} vanishes at the event horizon $r=r_+$
\beqa\label{mP}
m &=& -\frac{\Xi_{a_S}^2 \left(a_S^2 \left(2 Q-g^2 r_+^4\right)+g^2 r_+^6 \Xi_{a_S}+Q^2 \Xi_{a_S}^2+r_+^4\right)}{2 \left(a_S^2-r_+^2 \Xi_{a_S}\right)}
\equiv m_P
\label{mpope}
\eeqa
and  the fact that $\Omega(r_+) = \Omega_H$ is a parameter independent of  the charge $Q$ and $r_+$.  This yields
\beqa \label{aS}
a_S&=&\frac{2 L^2 r_+^4 \Omega_H}{r_+^4+L^2 \left(Q+r_+^2\right)+\sqrt{\left(L^2 \left(Q+r_+^2\right)+r_+^4\right)^2-4 L^4 Q r_+^4 \Omega_H^2}}   \equiv a_P
\label{apope}
\eeqa
upon replacing $m$ in \eqref{PopeKAdS}, where we note that the above expression has  a well-defined limit  as $Q\to 0$.

%\tcb{[why we should rename to $a_P$, $m_P$]}

Using \eqref{mP} and \eqref{aS} to replace $a_S$ with $a_P$ and $m$ with $m_P$, all metric functions in \eqref{PopeKAdS} depend only on the parameters $r_+$, $Q$, and $\Omega_H$.   We can expand the metric functions in any or all of these three parameters.  Despite the presence of terms linear in $Q$ in \eqref{PopeKAdS}, we find upon replacement of $a_S$ and $m$ using \eqref{mP} and \eqref{aS}
 that no metric function has any terms linear in $Q$ -- their series expansions begin with a $Q^2$ correction term to the metric functions in \eqref{KAdSS}.  We also find that the expansions begin with negative powers of $r_+$; this is consistent with the lack of an $r_+\to 0$ limit as there are no boson stars in the setting we are considering.\\

No exact solutions to the Einstein-AdS-Chern-Simons system of equations  (\ref{Eineq},\ref{Tab})
 for general values of $\lambda$ (including $\lambda=0$) and non-zero charge $Q$ are known. Here we aim to obtain such solutions perturbatively.
We use the above explicit expansions to confirm validity of our solutions that we obtained by directly solving equations of motion for supergravity value of CS coupling.

\section{Construction of the perturbative solution} \label{construction}

We expand the metric, gauge field in a power series in a small parameter $Q/l^2$ that is related to the physical charge (see \eqref{Chg} ). We then insert this expansion into the equations of motion, and expand latter in a power series in $Q/l^2$, although it turns out that the resultant Maxwell equations do not seem to be analytically solvable. Therefore we use the matching method (described below) to produce the solutions.

To this end, the starting point of our setup is a  KAdS black hole with arbitrary but small  $r_{+} \ll l$.  We introduce two dimensionless perturbative parameters: one associated with the horizon length, $r_+ /l$, and the other associated with the charge $Q/l^2$ of the gauge field.  Hence
we compute a  three-parameter family of solutions, performing a
 double expansion in these parameters analogous to that carried out for a scalar field \cite{Dias:2011at,Stotyn:2011ns}; our  third parameter is the horizon angular velocity $\Omega_H$.

The existence of the Chern-Simons term does not change the expression for the energy-momentum tensor, and so the expansion of this quantity begins at  quadratic order in $Q/l^2$. We therefore do not expect any terms linear in $Q$
in the metric functions, consistent with the exact solution \eqref{PopeKAdS} upon replacement of $a_S$ and $m$ using \eqref{mP} and \eqref{aS}.

Since the action remains invariant under the discrete symmetry
 \beqa
(Q,\lambda)\rightarrow (-Q,-\lambda)\label{dissym}
\eeqa
  changing only the sign of the charge (or alternatively the CS coupling) gives us another possible solution.
The symmetry \eqref{dissym} imputes certain properties to the perturbative expansion. For instance, even (odd) powers of Q in the metric field's expansions appear with even (odd) powers of $\lambda$. However, gauge field functions change sign under this transformation, $\ie$ even (odd) powers of Q accompany with odd (even) powers of $\lambda$. Therefore, we must take into account both odd and even powers in  $Q/l^2$  due to the presence of Chern-Simons term
\cite{Kunz:2010a04,Kunz:2010a07}.

As mentioned before, this  results in a set of ordinary differential equations that cannot be simultaneously solved everywhere in the spacetime, and so we shall employ a matched asymptotic expansion to deal with this situation.
This involves dividing the  exterior spacetime of the black hole into two regions: a near-region where $r_+\leq r\ll l$ and a far-region where $r\gg r_+$. In each of these regions, we can discard terms in the equations with sub-dominant contributions.
This yields in each region a coupled system of perturbative equations whose solutions are valid in different regions of approximation.
Provided $r_+/l \ll 1$, the far and near regions have an overlapping zone, $r_+\ll r \ll l$, and we  then match the set of independent parameters that are produced in each of the two regions \cite{Dias:2011at,Stotyn:2011ns,Minwalla:2010}.
At each order in $Q/l^2$ we have a set of linear differential equations, which we solve subject to the requirements of the normalizability of functions at infinity together with their regularity at the horizon.

In the next  section we describe details of the matching method. Then we present explicit results for our perturbative expansion in higher orders.  We first present the structure of the perturbative ansatz we employ in
both far and near field regions.

\subsection{Far Field Region} \label{far}

Let us first focus on the region $r \gg r_{+}$. In this region the background \reef{KAdSS} is a small perturbation about global AdS space.
As discussed above,   we start by performing a double expansion of our fields  in $\epsilon = Q/l^2$ and $r_+ / \ell$ as follows:
 \beqa
F^{out}(r,Q,r_+)=F^{out}_0(r,r_+)+F^{out}_2(r,r_+)\epsilon^2+\sum_{j=3}^n F^{out}_j(r,r_+) {\epsilon}^j
\label{eq:FarFieldExpansion}
\eeqa
where
\beqa
F^{out}_2(r,r_+)&=&F^{out}_{2,-2}(r)\left(\frac{l^2}{r_{+}^2}\right)+F^{out}_{2,0}(r)
+F^{out}_{2,2}(r)\left(\frac{r_{+}^2}{l^2}\right)+\ldots
\nonumber\\ F^{out}_j(r,r_+)&=&\sum_{i=-j}^p F^{out}_{j,2i+4}(r)\left(\frac{r_+}{l}\right)^{2 i+4}
\label{F-exp}
\eeqa
and we have explicitly set $F^{out}_1(r,r_+) = 0$ since (upon inspection of the field equations) terms linear in $Q$ are not present. The indices indicate the expansion order in $Q$ and $r_+$ respectively.

For the gauge field
\beqa
A^{out}(r,Q,r_+)=A^{out}_1(r,r_+)\epsilon+\sum_{j=2}^n A^{out}_j(r,r_+) {\epsilon}^j
\label{eq:FarFieldA}
\eeqa
with
\beqa
A^{out}_1(r,r_+)&=&A^{out}_{1,0}(r)
+A^{out}_{1,2}(r)\left(\frac{r_{+}^2}{l^2}\right)+\ldots
\quad \quad \quad A^{out}_j(r,r_+)=\sum_{i=-j}^p A^{out}_{j,2i+4}(r)\left(\frac{r_+}{l}\right)^{2 i+4}\nonumber
\eeqa
where $F^{out}=\{f^{out},g^{out},h^{out},\Omega^{out}\}$ is shorthand for each of the metric functions in (\ref{metric}) in the far-region   and ${A}^{out}$ denotes either component of the gauge field.
 The occurrence of successively
negative powers of $r_+$ with increasing powers of $Q$ in (\ref{F-exp}) will be seen to be necessary once we apply the method to match the outer series expansion with the inner one that necessarily includes negative powers  due to the form of the gauge field at leading order in $Q$.
The perturbative expansion then proceeds similarly to the boson star cases except that instead of our background being global AdS we take it to be the
 KAdS black hole, whose metric functions for  $Q=0$ are denoted by   $f_0(r),g_0(r),h_0(r)$ and $\Omega_0(r)$.

\subsection{Near Field Region} \label{near}

 For the near-field region $r_+\leq r \ll l$, the metric \eqref{metric} can be written as
\beqa
ds^2=\frac{r_+^2}{l^2}\left[-f\ g \ d\tau^2+\frac{dz^2}{f}+z^2\left[h\left(d\chi+\frac{\cos \theta}{2}d\phi-\frac{r_+}{l}\Omega d\tau\right)^2+\frac{1}{4}(d\theta^2+\sin^2\theta d\phi^2)\right]\right]\nonumber\\ \label{metricIn}
\eeqa
using the radial coordinate $z=r l/r_{+}$ and  $\tau=t l/r_+$. By applying the coordinate transformation, the rotation field emerges as $g_{\tau \psi} \propto \frac{r_+}{l} \Omega$. In the near region, as $r_+/l\ll 1$, it is suppressed relative to the other metric functions  and  can be constructed as a power series of $r_+/l$ around static configuration. This argument remains valid when we add charge gradually to the system and have another power series expansion in small $Q/l^2$.

For the KAdS metric, by applying this coordinate transformation on \eqref{KAdSS} we obtain
%where various fields are given by
\beqa
f&=&\frac{(L-z) (L+z)}{L^4 z^4} \left(\frac{L^6 \left(l^2+r_+^2\right)}{l^2 \left(1-r_+^2 \Omega_{H}^2\right)+r_+^2}-\left(l^2+z^2\right) \left(L^4+r_+^2 z^2\right)\right), \quad \quad g=\frac{1}{h} \nonumber\\
h &=&1-\frac{L^6 r_+^2 \Omega_{H}^2}{z^4 \left(l^2 \left(r_+^2 \Omega_{H}^2-1\right)-r_+^2\right)}, \quad \quad \Omega=\frac{L^4 \Omega_{H} \left(l^2+r_+^2\right)}{z^4 \left(l^2+r_+^2\right)+l^2 r_+^2 \Omega_{H}^2 \left(L^4-z^4\right)}\nonumber
\eeqa
for the various metric functions, where we have replaced $a_S$ and $m$ using \eqref{mP} and \eqref{aS} with $Q=0$.

The  horizon  is located at $z\simeq l$; hence $z/l \geq 1$, ensuring in turn that $z/l$ is larger than $r_+/l \ll 1$.  This yields a valid expansion in terms of the small horizon radius $r_+$.  Since the dependence of the
inner and outer solutions on $r_+$ differ in form, we  extend our expansions to include negative powers of $r_{+}$, writing
for the metric functions
 \beqa
F^{in}(z,Q,r_+)=F^{in}_0(z,r_+)+\sum_{j=2}^n F^{in}_j(z,r_+) \left(\frac{Q}{l^2}\right)^j
\qquad F^{in}_j(z,r_+)=\sum_{i=-j}^p F^{in}_{j,2i}(z)\left(\frac{r_+}{l}\right)^{2 i}
\label{FInexp}
\eeqa
where we study the above series up to order four in charge ($\ie$ $n=4$).  Based on the preceding  discussion, we use a similar expansion  $F^{in}=\{f^{in}, g^{in}, h^{in}, \Omega^{in}\}$ for the gravitational fields, where ``in" denotes these series expansions are used only in the near-region. The near-region fields are obtained by solving the equations of motion \eqref{Eineq} and imposing the horizon boundary conditions \eqref{bndry}.

This situation likewise holds for the gauge field, and so the time component of the gauge field is given by
\beqa
A_{\tau}^{in}(z,Q,r_+)={A_{\tau}^{in}}_1(z,r_+)\epsilon+\sum_{j=2}^n {A_{\tau}^{in}}_j(z,r_+) {\epsilon}^j\label{A0Inexp}
\eeqa
where
\beqa
{A_{\tau}^{in}}_1(z,r_+)&=&{A^{in}_{\tau}}_{1,-1}(z)\left(\frac{l}{r_{+}}\right)+{A^{in}_{\tau}}_{1,1}(z)\left(\frac{r_{+}}{l}\right)+{A^{in}_{\tau}}_{1,3}(z)\left(\frac{r_{+}^3}{l^3}\right)+\ldots\nonumber\\
{A_{\tau}^{in}}_j(z,r_+)&=&\sum_{i=-j}^p {A_{\tau}^{in}}_{j,2i+3}(r)\left(\frac{r_+}{l}\right)^{2 i+3}\nonumber
\eeqa
written in terms of the $\tau$ coordinate. The angular component of the gauge field does not include the time coordinate and so
\beqa
A_1^{in}(z,Q,r_+)=\sum_{j=1}^n {A_1^{in}}_j(z,r_+) {\epsilon}^j
\quad \quad \quad
{A_1^{in}}_j(z,r_+)=\sum_{i=-j}^p {A_1^{in}}_{j,2i+2}(r)\left(\frac{r_+}{l}\right)^{2 i+2}\label{A1Inexp}
\eeqa
retaining only even powers in $r_+$.  Note that in order to match the near-field with the far-field solutions  the  same time coordinate  $t$ must be used.

\subsection{Matching Procedure }\label{pert-result}

At leading order  in $Q/l^2$, $(n=1)$, we get a system of two linear differential equations from inserting the field expansion into the gauge field equations; it yields the expected $1/r^2$ behaviour for the zero component of the gauge field (chosen to be  independent of $r_+$). The angular component of the gauge field has the same radial function to desired higher order solutions in $r_+/l$ and decays as the boundary condition  \eqref{metbc} dictates.
At this order, gauge field does not react to the gravitational field and the background spacetime is the $(d=5)$ KAdS black hole introduced in \eqref{KAdSS}. From this analytic solution, the $n=0$ coefficients $F_0^{out}$ are known to any order $\cO(r_+^p/l^p)$. One only needs to insert the expressions of rotations parameter $a_M$ and mass-radius $r_M$ of the KAdS black hole \eqref{KAdSS} to gravitational functions in order to get series expansion in $r_+/l$. These expansions satisfy the gravitational field equations \eqref{Eineq} and the asymptotic boundary conditions \eqref{metbc} up to order $\cO(Q/l^2, r_+^p/l^p)$, as expected.

The near-region analysis is similar. At leading order in $Q$, we again have a linear perturbation problem. The gravitational coefficients $F_0^{in}$ are obtained from taking a series expansion in $r_+/l$ of the KAdS background up to the desired order. Gauge field coefficients ${A_1^{in}}_{1,2i+2}$ satisfy \eqref{Eineq} up to order $\cO(Q/l^2, r_+^p/l^p)$ and boundary condition in \eqref{bndry}.

\subsubsection{Matching region}

The far-region fields are valid for $r\gg r_+$ and the near ones for $r\ll l$. For small black holes we have $r_+\ll l$. The far and near region solutions overlap where $r_+\ll r \ll l$. Therefore this intersecting region implies matching conditions that fix the remaining integration constants to get solutions for the whole spacetime. The matching technique must be done at each order of $r_+/l$ for a given order of $Q/l^2$. In this procedure we match the near-region solution for large $r$ with far-region solution as $r$ becomes small. For this purpose, we take the Taylor expansion of \eqref{FInexp}, \eqref{A0Inexp}, \eqref{A1Inexp} for large $z$ and use the original radial coordinate $z=r l/r_+$, as well as the series expansion of \eqref{eq:FarFieldExpansion}, \eqref{eq:FarFieldA} for small-$r$. In employing this procedure, recall that for far-region solutions the coefficients of powers of $r_+/r$  must be either removed or matched with the same terms in the near region. Furthermore, terms behaving like $r/l$ in the inner region must either be eliminated or have counterparts in the far region solution.

Next we consider non-linear contributions $\cO(Q^n/l^{2 n}, r_+^p/l^p)$, $(n\geq 2)$,  where the field strength of the gauge field acts as a source in the Einstein equation, providing information about charged rotating black holes. We solve the coupled system of equations \eqref{Eineq} order by order in $(n, p)$. From the charge corrections of the Maxwell solutions at order $Q^n/l^{2 n}$, we proceed to $Q^{n+1}/l^{2 n+2}$ order to find the back-reaction of inserting extra charge on the gravitational fields.

%\tcb{We again use the far and near region analysis of the solutions and match them in the overlapping region, order by order like we described at the linear order. However, in higher orders, the perturbative procedure becomes tedious, due to occurrence of many fields and especially polylog terms that are not easy to manipulate.}

%\tcb{The next-leading-order is acquired by solving Maxwell equation up to $\cO(Q^2/l^4)$ as mentioned we fix the form of $a_0$ radially such that it does not include $r_+/l$ dependence. For angular component of gauge field we solved Maxwell equation for different orders in $r_+/l$, imposing boundary condition \eqref{metbc}, \eqref{bndry} that fixe one of two integration constants appearing  in each order in $r_+/l$ to make it physical.}
We have solved the system of equations (\ref{Eineq},\ref{Tab}) using the ansatz (\ref{metric},\ref{Aansatz}) to order $Q^4/l^8$, retaining
powers of $r_+$ as relevant.
We assume that $\cO(Q/l^2)\sim\cO(r_+/l)$. We managed to perform the perturbation method up to combined order $(n+p)\geq 5$ up to $Q^3/l^6$. However at order $Q^4/l^8$ the calculations become
extremely  cumbersome, and we did not obtain high enough corrections in $r_+/l$, to be compatible with our lower order results.  We therefore do not make use of them to obtain the physical quantities in section \ref{explor}.

Upon  solving the field equations, the  arbitrary constants for the near-field solution that appear are fixed  using the boundary conditions \eqref{bndry} at the horizon and by matching with the far-field solution. Likewise the constants in
the far-field solution are fixed using the boundary conditions \eqref{metbc}.
The results are very lengthy and so we present them in Appendix \ref{pert}.  In this subsection
we confine our remarks to some comments about our results.
We find for both the near-field and far-field expansions that the metric and gauge fields exhibit $1/r^n$
 radial behavior. This is due to the existence of the gauge field, and  is as expected for a charged black hole.
 Up to order $Q^3$, the solutions to the  Einstein equations  initially yield $\log$ and polylog terms that disappear in our final expressions after fixing constants from the boundary conditions; furthermore,
and the log and polylog terms appearing in higher order are removed for $\lambda=\lambda_{SG}$.
In general, to fix the constants of integration for $\Omega(r)$, we  used the fact that $\Omega_H$ is the final value for the horizon angular velocity (without the necessity of including contributions from Q and $r_+$).
It means the function $\Omega(r)$ is such that its corrections in Q vanish as $z\rightarrow l$.
 Only $\Omega (r)$ and the angular component of the gauge field depend on  odd powers of $\Omega_H$. When the sign of the azimuthal angular coordinate $\phi$ is changed, $\Omega_H\rightarrow -\Omega_H $;  we will see in section \ref{explor} that this  gives solutions with $J\rightarrow -J $.
 We note that $a_0^{in}$ in our approximation does not receive $Q^2$ corrections (see\eqref{a0inq2}).

\subsubsection{The Einstein-Maxwell  Case }\label{pertME}

One useful by-product of our results is that  for vanishing $\lambda$ we also obtain expansions for charged rotating AdS black holes of equal angular momenta
in Einstein-Maxwell theory. No analytic solutions are known for this case.  Only even powers of $Q$ in any metric function and only odd powers of $Q$ in the gauge field components appear, and
their explicit expressions  can be obtained by setting $\lambda=0$ in our results in the appendix.  It is computationally much easier to go to even higher orders in $Q$ but we have not included these expressions here.

\section{Conserved Quantities and Thermodynamics} \label{thermo}

The metric   \eqref{metric}
has five independent Killing vector fields: $\partial_t$, $\partial_{\psi}$ and the three rotations of $S^2$. However the presence of the gauge field yields a reduced level of symmetry, with  the Killing vector given by
\beqa
\zeta &=& \partial_t+\Omega_H \partial_{\psi}\nonumber
\eeqa
That is null on and orthogonal to the event horizon $r=r_+$, which is therefore a Killing horizon. The horizon electrostatic potential $\Phi_H$ is given by
\beqa
\Phi_H &=& (a_0+\Omega_H a_1)|_{r=r_+}\nonumber
\eeqa
The expression for finding the electric charge is the Gaussian integral\\
\beqa
Q_E&=&\frac{1}{16\pi}\int_{S^3_{\infty}}(\star F-F\wedge A/\sqrt{3})\nonumber
\eeqa
%\tcr{\bf [You had the above equation commented-out in the previous version.  Why?]}
over the $S^3$ at infinity,
where in 5 dimensions, the CS term will vanish when $r\rightarrow \infty$ and at the end only
the non-zero term in $a_0^{out}$ expansion in \eqref{a0out}
% $C_{a_0}$ in (?)
  gives a contribution.

We use the Ashtekar-Das formalism \cite{Ashtekar:2000,Das:2000cu} %\tcr{\bf [Fix reference]}
to calculate conserved quantities associated with the symmetries coming from asymptotic Killing vectors.  For an asymptotic conformal Killing vector $\xi$, the corresponding conserved charge $Q_{\xi}$ associated with this symmetry is
\beqa
Q_{\xi}=\pm \frac{d-3}{8\pi}\int_{\Sigma}\tilde{\varepsilon}_{a b}\xi^a t^b d\Sigma,\nonumber
\eeqa
where $\pm$ denotes a timelike/spacelike conformal Killing vector, respectively, and
$\tilde{\varepsilon}_{a b}=l^2 \tilde{K}_{a b c d} n^b n^d$ is the electric part of the conformal Weyl tensor, where
$\tilde{K}_{a b c d}=\tilde{\Lambda}^{3-d}\tilde{C}_{a b c d}|_{\tilde{\Lambda}=0}$ and $\tilde{\Lambda}=1/r$ is the conformal factor.
For $\xi=(\partial_t,\partial_{\psi})$ we obtain
\beqa
E=\frac{\pi l^2 }{8}(4C_h-3C_f),~~~~ J=\frac{\pi l^3 C_{\Omega}}{2}\nonumber
\eeqa
for the energy and angular momentum respectively,
where the constants $C_f, C_h, C_{\Omega}$ are given by the boundary conditions \eqref{metbc}. % \tcr{\bf [Fix eq. reference]}

The temperature   is specified by the surface gravity as $T_H=\sqrt{-(\nabla \zeta)^2|_{r_+}}/(2\sqrt{2}\pi)$ and the entropy   is given by the area of black hole divided by 4. Hence
\beqa
T_H=\frac{f^\prime\sqrt{g}}{4\pi}|_{r_+},~~~~ S=\frac{1}{4}A_{d-2}r_+^{d-2}\sqrt{h(r_+)}\nonumber
\eeqa
where $A_{d-2}$ is the area of $(d-2)$-sphere.

Using the   perturbative  solutions in appendix \ref{pert} we find
%Electric charge
\beqa
Q_E&=&\frac{\sqrt{3}}{4}  \pi  l^2 \left(\frac{Q}{l^2}\right)
\label{Chg}
\eeqa
 %Energy
\beqa
E&=&\frac{\pi  l^2}{4}\bigg[\frac{3}{2} \left(\frac{r_+^2}{l^2}\right)+\frac{3}{2} \left(l^2 \Omega_H^2+1\right)\left(\frac{r_+^4}{l^4}\right)+\frac{l^2 \Omega_H^2}{2} \left(3 l^2 \Omega_H^2+1\right)\left(\frac{r_+^6}{l^6}\right) +\frac{ l^2 \Omega_H^2}{2} \left(l^2 \Omega_H^2 -1\right)\nonumber\\
&&\left.\times\left(3 l^2 \Omega_H^2+1\right)\left(\frac{r_+^8}{l^8}\right)+\cO\left(\frac{r_+^{10}}{l^{10}}\right) +\left(\frac{Q^2}{l^4}\right)\bigg(\frac{3}{2}\left(\frac{l^2}{r_+^2}\right)-\frac{3 l^4 \Omega_H^4 }{2} \left(\frac{r_+^2}{l^2}\right) -\frac{l^2 \Omega_H^2}{2}  \big(l^2\right.\nonumber\\
&&\left.\times \Omega_H^2\left(3 l^2 \Omega_H^2-8\right)-1\big)\left(\frac{r_+^4}{l^4}\right)
+\cO\left(\frac{r_+^{6}}{l^{6}}\right)\bigg)+\left(\frac{Q^3}{l^6}\right)\bigg(6 \sqrt{3} \lambda  l^4 \Omega_H^4-2 \sqrt{3} \lambda  l^2 \Omega_H^2 \big(11\right.\nonumber\\
&&\left.\times l^2 \Omega_H^2+1\big)\left(\frac{r_+^2}{l^2}\right)
+\cO\left(\frac{r_+^{4}}{l^{4}}\right)\bigg)+\left(\frac{Q^4}{l^8}\right)\bigg(\frac{1}{3} \left(1-12 \lambda ^2\right) l^2 \Omega_H^2 \left(\frac{l^4}{r_+^4}\right)+\cO\left(\frac{l^{2}}{r_+^{2}}\right)\bigg)
\bigg]\right.\nonumber\\
\label{Energy}
\eeqa
\beqa
J&=&\frac{\pi  l^3}{2}\bigg[l \Omega_H\left(\frac{r_+^4}{l^4}\right) +l^3 \Omega_H^3\left(\frac{r_+^6}{l^6}\right) +l^3 \Omega_H^3 \left(l^2 \Omega_H^2-1\right)\left(\frac{r_+^8}{l^8}\right) +\cO\left(\frac{r_+^{10}}{l^{10}}\right) +\left(\frac{Q^2}{l^4}\right)\nonumber\\
&&\left.\times\bigg( l \Omega_H-l \Omega_H \left(l^2 \Omega_H^2+1\right)\left(\frac{r_+^2}{l^2}\right) -l \Omega_H \left(l^2 \Omega_H^2 \left(l^2 \Omega_H^2-3\right)-1\right)\left(\frac{r_+^4}{l^4}\right) \right.\nonumber\\
&&\left.+\cO\left(\frac{r_+^{6}}{l^{6}}\right)\bigg)
+\left(\frac{Q^3}{l^6}\right)\bigg(-2 \sqrt{3} \lambda l \Omega_H\left(\frac{l^2}{r_+^2}\right) +2 \sqrt{3} \lambda  l \Omega_H \left(3 l^2 \Omega_H^2+2\right)-2 \sqrt{3} \lambda  l \Omega_H \right.\nonumber\\
&&\left.\times\left(11 l^2 \Omega_H^2+3\right)\left(\frac{r_+^2}{l^2}\right)+\cO\left(\frac{r_+^{4}}{l^{4}}\right)\bigg)
+\left(\frac{Q^4}{l^8}\right)\left(12 \lambda ^2 l \Omega_H\left(\frac{l^4}{r_+^4}\right)+\cO\left(\frac{l^{2}}{r_+^{2}}\right)\right)
\bigg]\right.
\label{Angmom}
\\
S&=&\frac{\pi ^2 r_+^3}{2}\bigg[1+\frac{l^2 \Omega_H^2}{2}  \left(\frac{r_+^2}{l^2}\right)+\frac{l^2 \Omega_H^2}{8}  \left(3 l^2 \Omega_H^2-4\right)\left(\frac{r_+^4}{l^4}\right) +\cO\left(\frac{r_+^{6}}{l^{6}}\right) +\left(\frac{Q^2}{l^4}\right)\bigg(-\frac{l^2 \Omega_H^2}{2}\nonumber\\
&&\left.\times \left(\frac{l^2}{r_+^2}\right)
-\frac{l^2 \Omega_H^2}{4}  \left(l^2 \Omega_H^2-6\right)+\cO\left(\frac{r_+^{2}}{l^{2}}\right)\bigg)
+\left(\frac{Q^3}{l^6}\right)\bigg(2 \sqrt{3} \lambda  l^2 \Omega_H^2 \left(\frac{l^4}{r_+^4}\right)-\sqrt{3} \lambda  l^2 \right.\nonumber\\
&&\left.\times\Omega_H^2 \left(l^2 \Omega_H^2+8\right)\left(\frac{l^2}{r_+^2}\right)+\cO(1)\bigg)
+\left(\frac{Q^4}{l^8}\right)
\left(\cO\left(\frac{l^{6}}{r_+^{6}}\right)\right)
\bigg]\right.
\label{Entropy}
\\
T_H&=&\frac{1}{\pi  r_+}\bigg[\frac{1}{2}+\left(1-\frac{3 l^2 \Omega_H^2}{4} \right)\left(\frac{r_+^2}{l^2}\right) +\frac{l^2 \Omega_H^2}{16}  \left(4-5 l^2 \Omega_H^2\right)\left(\frac{r_+^4}{l^4}\right) -\frac{l^2 \Omega_H^2}{32} \Big(l^2 \Omega_H^2 \big(7 l^2 \Omega_H^2\nonumber\\
&&\left.-16\big)
+8\Big)\left(\frac{r_+^6}{l^6}\right) -\frac{l^2 \Omega_H^2}{256}  \left(l^2 \Omega_H^2 \left(l^2 \Omega_H^2 \left(45 l^2 \Omega_H^2-152\right)+176\right)-64\right)\left(\frac{r_+^8}{l^8}\right) \right.\nonumber\\
&&\left. +\cO\left(\frac{r_+^{10}}{l^{10}}\right) +\left(\frac{Q^2}{l^4}\right)\bigg(-\frac{1}{2}\left(\frac{l^4}{r_+^4}\right) +\frac{l^2 \Omega_H^2}{2}\left(\frac{l^2}{r_+^2}\right) +\frac{l^2 \Omega_H^2}{16}  \left(3 l^2 \Omega_H^2-8\right)+\frac{l^2 \Omega_H^2}{8}  \right.\nonumber\\
&&\left.\times\left(l^2 \Omega_H^2-6\right)
\left(l^2 \Omega_H^2-1\right)\left(\frac{r_+^2}{l^2}\right) +\frac{5}{256} l^2 \Omega_H^2 \Big(l^2 \Omega_H^2 \left(l^2 \Omega_H^2 \left(5 l^2 \Omega_H^2-32\right)+112\right)\right.\nonumber\\
&&\left.-64\Big)\left(\frac{r_+^4}{l^4}\right)
+\cO\left(\frac{r_+^{6}}{l^{6}}\right)\bigg)
+\left(\frac{Q^3}{l^6}\right)\bigg(-\sqrt{3} \lambda  l^2 \Omega_H^2\left(\frac{l^4}{r_+^4}\right) +\frac{\sqrt{3}}{2}  \lambda  l^2 \Omega_H^2 \left(l^2 \Omega_H^2+4\right)\right.\nonumber\\
&&\left.\times\left(\frac{l^2}{r_+^2}\right) +\frac{\sqrt{3}}{8} \lambda  l^2 \Omega_H^2 \Big(l^2 \Omega_H^2 \left(l^2 \Omega_H^2+4\right)-32\Big)+\frac{\sqrt{3}}{16}  \lambda  l^2 \Omega_H^2 \Big(l^2 \Omega_H^2 \big(l^2 \Omega_H^2 \big(l^2 \Omega_H^2\right.\nonumber\\
&&\left.-24\big)-72\big)+128\Big)\left(\frac{r_+^2}{l^2}\right)
+\cO\left(\frac{r_+^{4}}{l^{4}}\right)\bigg)
+\left(\frac{Q^4}{l^8}\right)\bigg(\frac{1}{18} \left(192 \lambda ^2-7\right) l^2 \Omega_H^2 \left(\frac{l^6}{r_+^6}\right)\right.\nonumber\\
&&\left. +\frac{1}{144} l^2 \Omega_H^2 \Big(-24 \lambda ^2 \left(133 l^2 \Omega_H^2+4 \pi ^2+146\right)
+77 l^2 \Omega_H^2+8 \pi ^2+76\Big)\left(\frac{l^4}{r_+^4}\right)+\cO\left(\frac{l^{2}}{r_+^{2}}\right)\bigg)
\bigg]\right.\nonumber\\
\label{Temp}
\\
{\Phi}_H&=&\frac{1}{r_+^2}\bigg[\left(\frac{Q}{l^2}\right) \bigg(\sqrt{3} l^2 \left(1+l^2 \Omega_H^2 \left(-\left(\frac{r_+^2}{l^2}\right)+\left(\frac{r_+^4}{l^4}\right)-\left(\frac{r_+^6}{l^6}\right)+\left(\frac{r_+^8}{l^8}\right)
-\left(\frac{r_+^{10}}{l^{10}}\right)\right)\right)\nonumber\\
&&\left.
+\cO\left(\frac{r_+^{12}}{l^{12}}\right)\bigg)
+\left(\frac{Q^2}{l^4}\right)\bigg(6 \lambda  l^4 \Omega_H^2-6 \lambda  l^4 \Omega_H^2 \left(l^2 \Omega_H^2+2\right)\left(\frac{r_+^2}{l^2}\right) +18 \lambda  l^4 \Omega_H^2 \big(l^2 \Omega_H^2\right.\nonumber\\
&&\left.+1\big)\left(\frac{r_+^4}{l^4}\right) -12 \lambda  l^4 \Omega_H^2 \left(3 l^2 \Omega_H^2+2\right)\left(\frac{r_+^6}{l^6}\right)+\cO\left(\frac{r_+^{8}}{l^{8}}\right)\bigg)
+\left(\frac{Q^3}{l^6}\right)
\bigg(\frac{l^4 \Omega_H^2}{\sqrt{3}}\big(1\right.\nonumber
\eeqa
\beqa
&&\left.-48 \lambda ^2\big) \left(\frac{l^2}{r_+^2}\right)+\frac{l^4 \Omega_H^2 }{\sqrt{3}} \left(144 \lambda ^2+\left(120 \lambda ^2-1\right) l^2 \Omega_H^2-3\right)
-\frac{2 l^4 \Omega_H^2}{\sqrt{3}}  \Big(12 \lambda ^2 \left(l^2 \Omega_H^2+6\right) \right.\nonumber
\\
&&\left.\times\big(3 l^2 \Omega_H^2+2\big)-2 l^2 \Omega_H^2-3\Big)
\left(\frac{r_+^2}{l^2}\right) +\cO\left(\frac{r_+^{4}}{l^{4}}\right)
\bigg)
+\left(\frac{Q^4}{l^8}\right)\bigg(4 \lambda  \left(30 \lambda ^2-1\right) l^4 \Omega_H^2\left(\frac{l^4}{r_+^4}\right) \right.\nonumber\\
&&\left.+2 \lambda  l^4 \Omega_H^2 \left(-240 \lambda ^2+\left(7-300 \lambda ^2\right) l^2 \Omega_H^2+8\right)
\left(\frac{l^2}{r_+^2}\right)+\cO(1)\bigg)\bigg]\right.
\label{Pot}
\eeqa
It is straightforward (but tedious) to check (up to the orders in our expansion that we have computed) that these quantities  satisfy the first law of thermodynamics
\beqa
d E=\Omega_H d J+T_H d S+Q_E d \Phi_H\nonumber
\eeqa
where the differentials can be computed by regarding $J$, $S$, and $\Phi_H$ as functions of the parameters $Q$, $r_+$ and $\Omega_H$.

\subsection{Further Exploration} \label{explor}

In this section we study relationships between the various  physical quantities associated with our solutions (rescaled by appropriate powers of $l$ to be dimensionless) up to order $Q^3/l^6$,
We define $\alpha = 2\sqrt{3}\lambda$ so that $\alpha=1$ is the supergravity solution \eqref{PopeKAdS}. As noted above, for $\alpha\neq 0$,  the  $Q \rightarrow -Q$ symmetry of Einstein-Maxwell theory is broken and so we need to consider positive and negative charges separately.

In plotting our results we impose several physical constraints.  First, we require non-negative values for $E$, $S$, and the temperature $T$.  The effect of this is that $|\Omega_H|$ is bounded by some maximal value dependent on the other parameters in the solution.  Furthermore, noting that
 \eqref{Energy} and \eqref{Entropy} are polynomials in $\Omega_H$, the requirement  that $E>0$
 is easily satisfied for small $\Omega_H$ since the leading term is positive, whereas the largest power
 ($\Omega^6_H$ to the order we are working)   implies that  $r_+^2>Q$.

\begin{figure*}[h]
\centering
\begin{tabular}{cc}
\includegraphics[scale=.35]{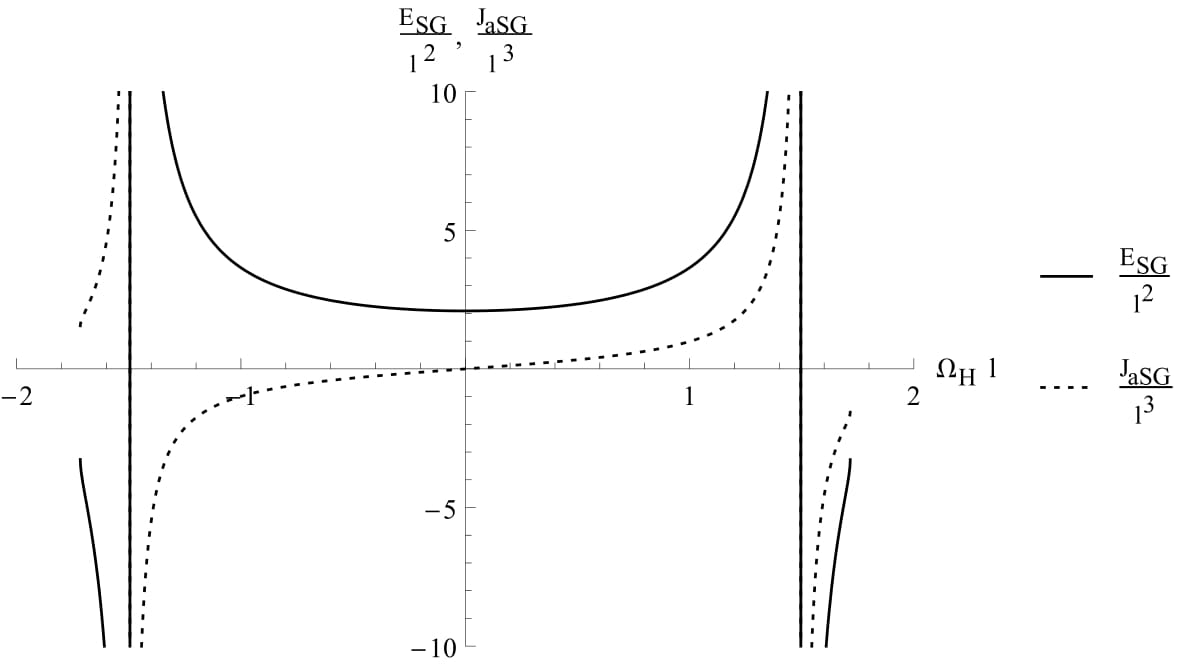}
\\
\end{tabular}
\caption{The exhibition of scaled $E_{SG}$ and ${J_a}_{SG}$ in terms of $\Omega_H l$}\label{EPJP}
\end{figure*}

All plots are for quantities computed to order $Q^3/l^6$ unless otherwise stated.  We
 can check the accuracy of our approximation by  comparing it to the physical quantities
associated with the exact solution \eqref{PopeKAdS} with $\alpha=1$.
The formulae for $E$ and $J$ for the supergravity solution are \cite{Pope:2005}
\beqa
E_{SG}&=&\frac{\pi  \left(4 a_S^2 g^2 Q \Xi_S^2-m (\Xi_S-4)\right)}{4 \Xi_S^3}\nonumber
\\
{J_a}_{SG}&=&\frac{\pi a_S  \left(Q \Xi_S^2 \left(a_S^2 g^2+1\right)+2  m\right)}{4 \Xi_S^3}\nonumber
\eeqa
where ${J_a}_{SG}$ denotes the angular momentum along one of two possible rotation planes in $d=5$.
Note that this is half the total angular momentum.
After substituting different parameters from \eqref{mpope} and \eqref{apope} we find a value for $\Omega_H$ where $E$ and $J$ in above formulae diverge; explicitly it is
\beqa
{\Omega_H^2}_{max}=\frac{r_+^2+1}{r_+^2}\nonumber
\eeqa
and for larger values $\Omega_H > {\Omega_H}_{max}$, $E_{SG}$ becomes negative.
For the specific choice of $r_+/l=.9 $ the upper bound for $\Omega_H l$ is $1.49485$.   We illustrate this situation in figure \ref{EPJP}.
%\graphicspath{ {E:/mozhganpro1/newpaper/newplots/} }
\begin{figure*}[h]
\centering
\begin{tabular}{cc}
\includegraphics[scale=.19]{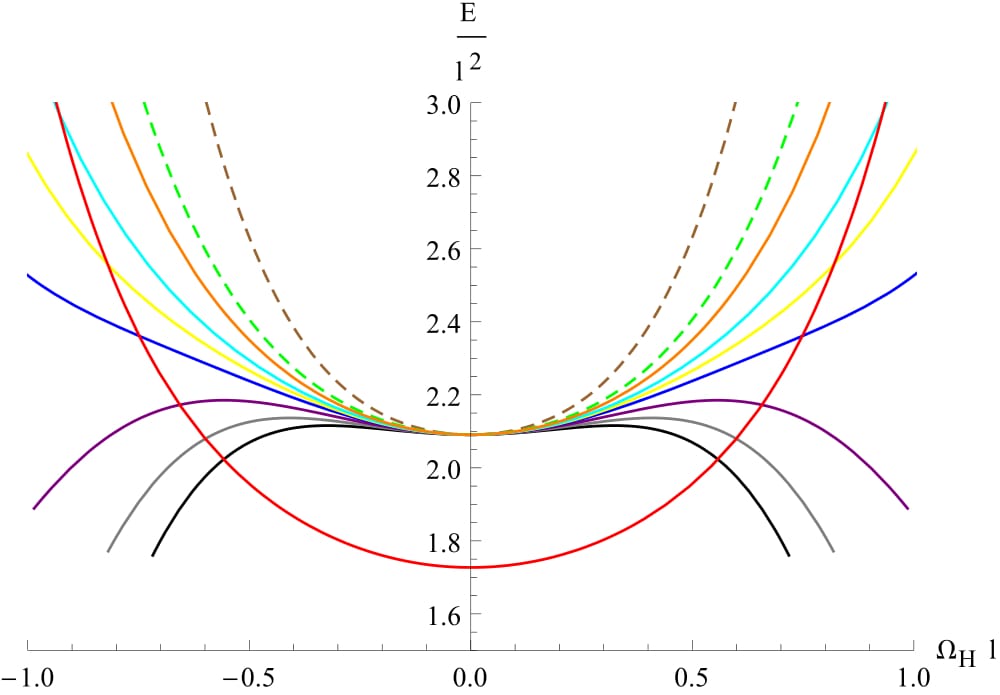}&
\includegraphics[scale=.19]{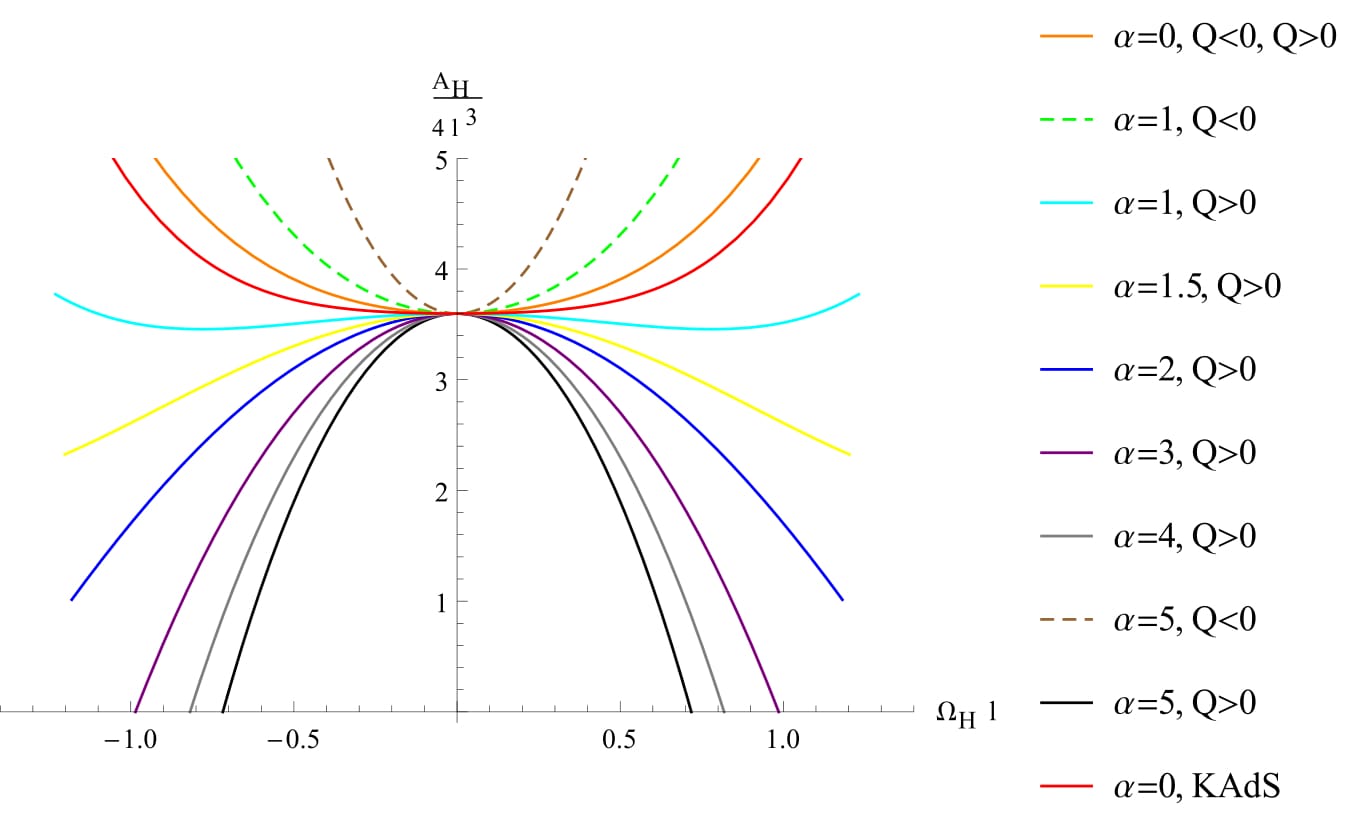}
\\
\end{tabular}
\caption{Energy (left) and entropy (right) versus  horizon angular velocity for $Q/l^2=0.5$, $r_+/l=0.9$, $\alpha =0, \cdots, 5$.  When  $\Omega_H=0$ all curves intersect
at $E/l^2 = 2.09082$ (a) and  $A_H/4 l^3 = 3.59747$ (b),
which corresponds to the RN-AdS solution.  For comparison the $Q=0$, $\alpha=0$ KAdS solution is also illustrated.  Curves for positive and negative charges are presented by solid and dashed lines respectively.}
\label{EWSW}
\end{figure*}

We have numerically checked that there is a broad region of $(Q,r_+,\Omega_H)$ parameter space
for which our results for the physical parameters of the black hole closely approximate those of the exact supergravity solution  \cite{Pope:2005}, for which $\alpha=1$.  All numerical values that appear in our plots are within this range. In subsequent plots we include the $\alpha=1$ case for which exact results are known.

We plot in fig. \ref{EWSW} the energy (a) and entropy (b) as functions of $\Omega_H$ for fixed
$Q=0.5 l^2$ and $r_+=0.9 l$.  The energy is initially an increasing function of $\Omega_H$ for all values of $\alpha$.  However at some value of  $\alpha >2$ the energy is maximized at some intermediate value
$\Omega/l^2_H < 1$, after which it is a decreasing function of $\Omega_H$. The entropy is likewise an increasing function of $\Omega_H$ for  $\alpha < 1$, but for larger values of $\alpha$ it monotonically decreases.
The black hole shrinks in size as its horizon velocity gets larger,
and for  $\alpha > 2$ we find that the entropy vanishes for values of $\Omega/l^2_H < 1$.
For both $E$ and $S$ we find for increasing $Q$ that the `fan' of lines spreads out more.
 For smaller values of Q different curves overlap or change slowly.

%\graphicspath{ {E:/mozhganpro1/newpaper/plotq3eps/} }
\begin{figure}[h]
\includegraphics[width=\textwidth]{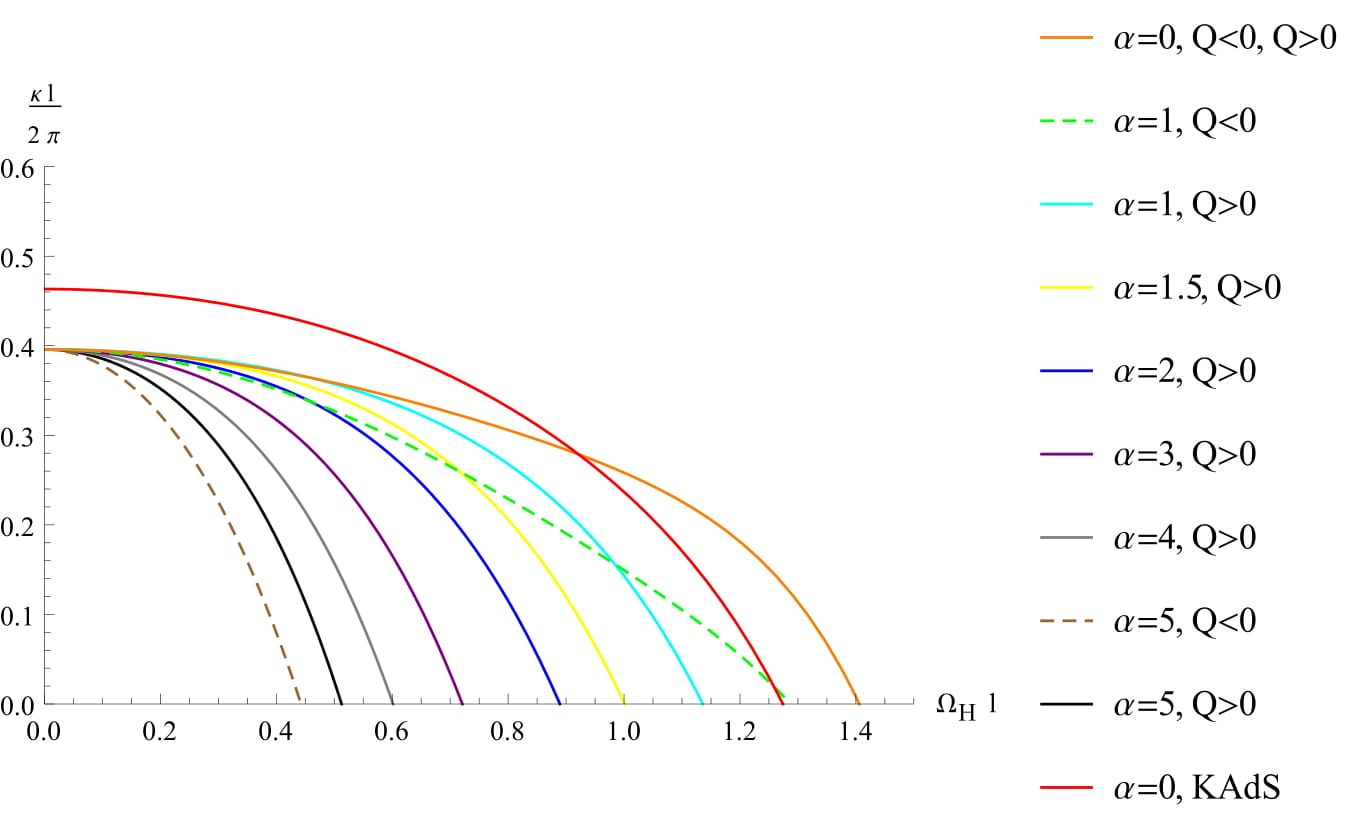}
\caption{ The temperature versus the horizon angular velocity. We choose $Q=0.5 l^2$, $r_+=0.9 l$ and $\alpha=0, \cdots 5$.
The intersection point at $\Omega_H=0$ is at 0.395935, corresponding to the RN-AdS solution.
%The intersection for KAdS solution where it tends to the nonrotating solution happens at 0.463318.
For comparison the $Q=0$, $\alpha=0$ KAdS solution is also illustrated.  Curves for positive and negative charges are presented by solid and dashed lines respectively.}
\label{TW}
\end{figure}

We plot in Fig. \ref{TW} temperature (or surface gravity $\kappa$) as a function of $\Omega_H$, retaining terms up to $Q^4/l^8$.
All curves end at an extremal solution with vanishing surface gravity and for all values of $\alpha$, $T$ is decreasing with horizon angular velocity.   As $\alpha$ increases the magnitude of the negative slope becomes increasingly large with increasing $\Omega_H$ for both signs of $Q$, with the intersection
point at extremality occurring for smaller values of $\Omega_H$.

 %\graphicspath{ {E:/mozhganpro1/newpaper/plotq3eps/} }
\begin{figure*}[h]
\centering
\begin{tabular}{cc}
\includegraphics[scale=.25]{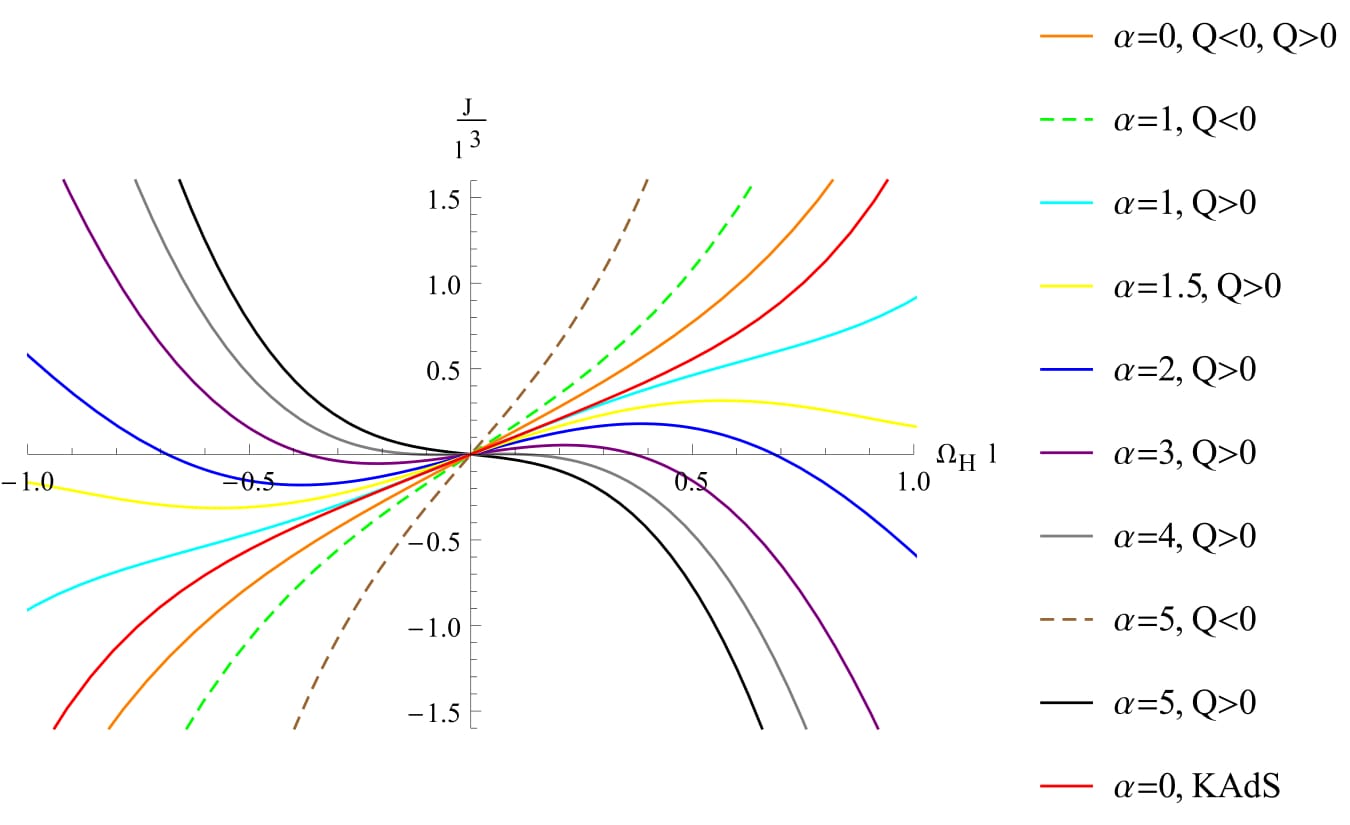}
 \\
\end{tabular}
\caption{The angular momentum versus the horizon angular velocity.
Curves for positive and negative charges are presented by solid and dashed lines respectively.}
\label{JWSJ}
\end{figure*}
%1-
%F(A-J)% presents the horizon area $A_H$ versus the J as and the corresponding curve for KAdS solution.

Figure \ref{JWSJ} contains a plot  of   the  angular momentum $J$ as a function of the angular velocity of the horizon.  For small values of $\alpha < 1$ the
angular momentum increases with increasing $\Omega_H$.  Increasing $\alpha$ further, another threshold is encountered marking the onset of a rotational instability, above  which there is a  maximal value of $J$ attainable for any $\Omega_H$.
    For values of $\alpha$ above this second threshold, there appear for a given $J$ two distinct solutions  that have the same $Q$ and  $r_+$ but different horizon angular velocities.
In particular there exist $J=0$ solutions having
    the same $Q$ and $r_+$ but differing angular velocities (and, we find,  horizon areas).
      For  even larger values
of $\alpha$, counterrotating black holes appear.  These are black holes whose angular momenta are opposite in sign to the angular velocity of their horizon \cite{Kleihaus:2004}.
For the representative choice $\Omega_H l=0.717879$,  the critical values for $\alpha$ at the two thresholds mentioned above are respective $1.35182$ and $1.93561$; in general these depend on the value of $\Omega_H$.

\begin{figure*}[h]
\centering
\begin{tabular}{cc}
\includegraphics[scale=.25]{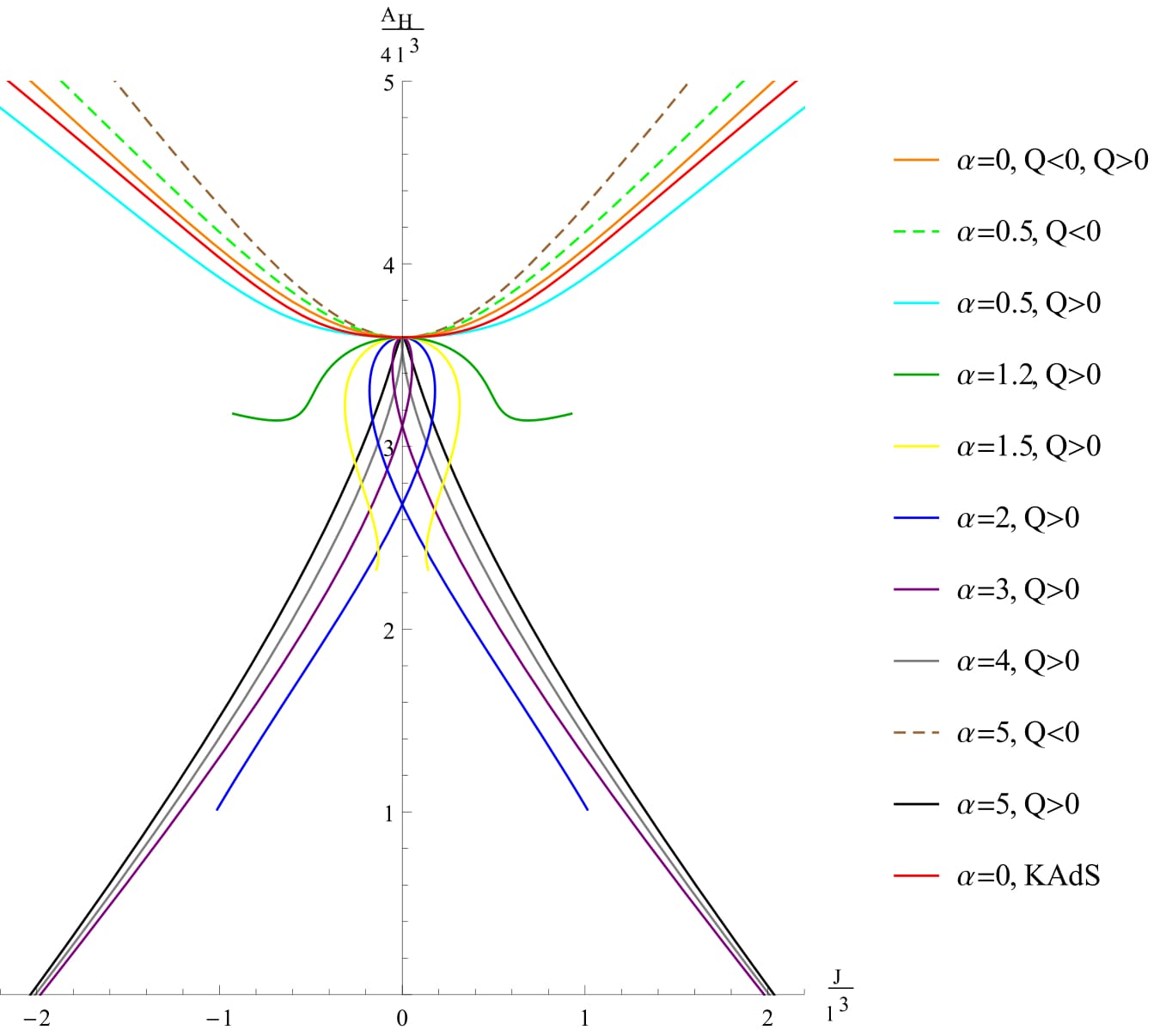}
 \\
\end{tabular}
\caption{Horizon area versus the angular momentum.
Curves for positive and negative charges are presented by solid and dashed lines respectively.}
\label{AWSJ}
\end{figure*}

In figure \ref{AWSJ} we plot horizon area (or entropy) as a function of angular momentum. We see that the horizon area increases with increasing $J$  from its RN-AdS value  at $J=0$ for $\alpha<1$.  For  $1  < \alpha < 3$  there is threshold above which the area is a decreasing function of increasing $J$.
%In between these values
%In the large $\alpha>3$ regime,  as $J$ increases the area decreases:
The   black holes become more `squashed' as $J$ gets larger. We find that there are two possible values of the area for one value of $J$.  We also find    in this regime there exist black holes of the same area with two equal angular momenta with opposite signs.

For the $\alpha \gtrsim 3$
we see from figure \ref{AWSJ}   that  $A_H$ goes to zero at some finite value of $J$.   This phenomenon has been seen previously for extremal charged rotating black holes in Chern-Simons theory that are asymptotically flat, and is indicative of a rotational instability:  for increasing $\alpha$, instead of rotating faster, black holes start to quench \cite{Kunz:2015}.

 The occurrence of  solutions with   zero total angular momentum is due to the cancellation between the horizon angular momentum and the angular momentum of the Maxwell field outside the horizon \cite{Kunz:2005,Kunz:2015}. For $\alpha<2$, the $J=0$ solutions   must be static but for $\alpha\geq 2$ they are  stationary  (non-zero $\Omega_H$) as appears in \cite{Straumann:1997}. This is illustrated in  fig \ref{JWSJ}. We also see that
  for sufficiently large  $\alpha$   these   solutions are not present.

The above results qualitatively hold for all  small positive charges $Q>0$.  For $Q<0$ all solutions have physical quantities
whose qualitative dependence on $\Omega_H$  (or alternatively $J$) is similar to that for
$Q>0$ and  $\alpha<1$; for example for all $Q<0$,  horizon area increases as $J$ increases.
Comparing Figures in \ref{JWSJ} we see that
$\Omega_H$ is also well-behaved while $A_H$ shrinks to zero.
%When $\alpha\geq 2$, we get solutions with $J=0$ as appears in \cite{Straumann:1997}.

We also exhibit in figures \ref{JWSJ} and \ref{AWSJ} the effects of changing the signs of $\Omega_H$ and $Q$.
For a given solution, changing $\Omega_H\rightarrow -\Omega_H$ yields a solution identical
to one with the same parameters except for $J\rightarrow -J$. %as shown in
%fig. \ref{JWSJ}(a).
Counterrotation is illustrated quite clearly in Fig. \ref{JWSJ} for $\alpha=5$: we see
that for $Q>0$ the sign of $\Omega_H$ is always opposite to that of $J$ for sufficently large
$\alpha$.  Indeed,
 (as noted previously) the only symmetry associated with changing $Q\to -Q$ is that of simultaneously changing $\lambda\to -\lambda$.  Hence for a fixed $\lambda$
solutions with $Q<0$ behave very differently from those with $Q>0$, as illustrated in fig. \ref{JWSJ}, which illustrates the $J\to -J$ symmetry, but not $Q\to -Q$, the distinction in the latter case growing more pronounced with increasing $\alpha$.

%%%%%%%%%%%%%%%%%%%%%%%%%%%%%%%%%%%%%%%%%%%%%%%%%%%%%%%%5
%\graphicspath{ {E:/mozhganpro1/newpaper/plotq3eps/} }
%\graphicspath{ {E:/mozhganpro1/newpaper/newplots/} }
\begin{figure*}[h]
\centering
\begin{tabular}{cc}
\includegraphics[scale=.3]{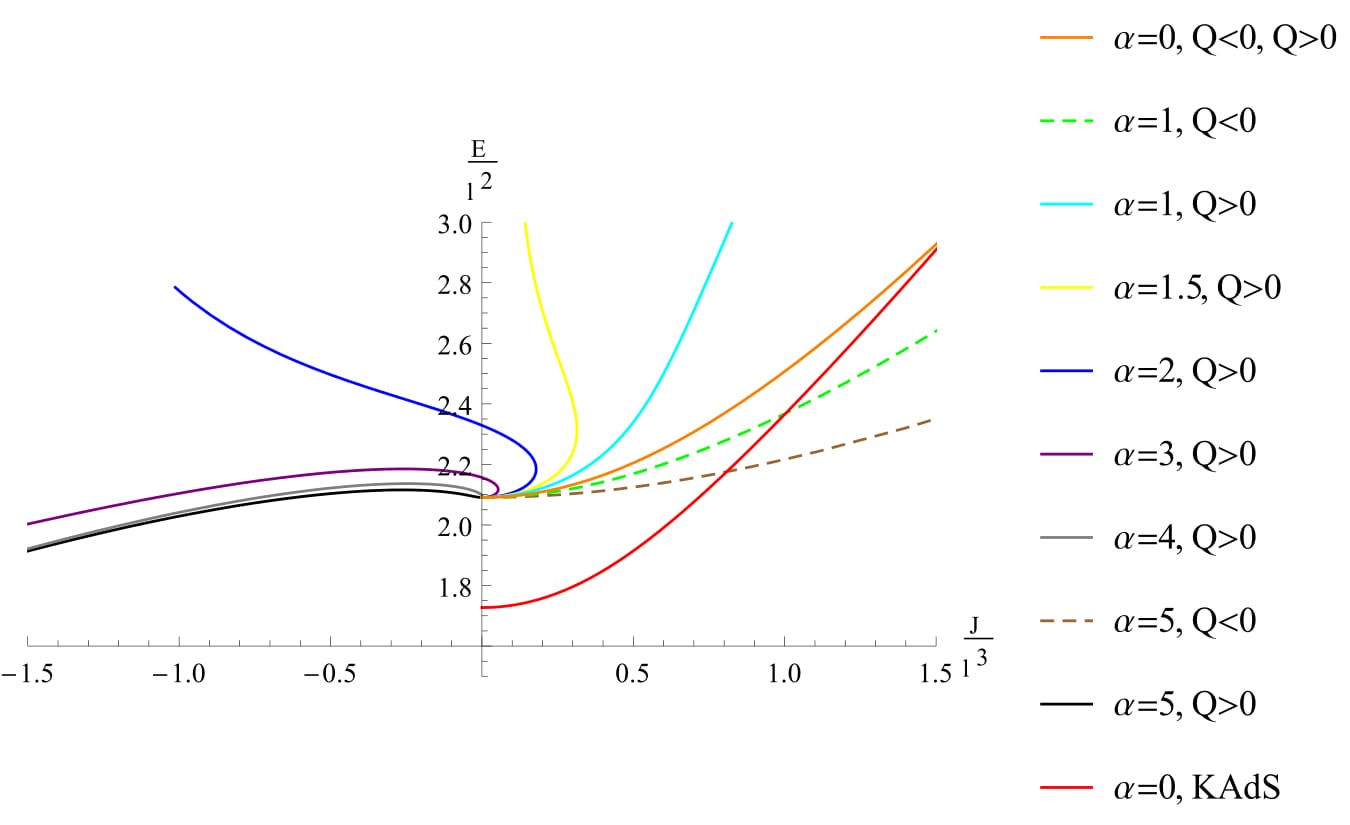}\\
\end{tabular}
\caption{Energy versus the angular momentum for   charge $|Q|/l^2= 0.5 $ and $r_+/l=0.9 $. The  intersection with the energy axis at $E/l^2
= 2.09082$  for $J=0$ is the RN-AdS solution.   For completeness we include the $Q=0$ KAdS solution. Curves for positive and negative charges are presented by solid and dashed lines respectively.  These curves all reflect about the $J=0$ axis for $\Omega_H < 0$; we have not drawn the associated graph.
}\label{EJ;alpha,Q}
\end{figure*}

To further illustrate the effects of the sign of $Q$ we plot in
fig. \ref{EJ;alpha,Q}  energy as a function of the total angular momentum $J$ for
$Q>0$ and $Q<0$ for various values of $\alpha$. For $Q<0$ we see that $E$ is an
increasing function of $J$ for all values of $\alpha$.  However for $Q>0$ we see the same
threshold values of $\alpha$ arise, with counter-rotation clearly visible for $\alpha=4,5$ (where the $J=0$ solutions disappear)  and
double-valued behaviour for intermediate values  $\alpha=2,3$.

%In Fig. \ref{EJ;alpha,Q}(a) especially by choosing $\Omega_H$, for value of $\alpha>$, we get $J=0$ solutions and for larger values of $\alpha$, J becomes negative.\\
We also see from figures \ref{JWSJ} and \ref{EJ;alpha,Q}  that the black hole solutions are not unique.
 As previously noted, there is a critical value $\alpha_{cr}$ above which there are $J=0$ ($\Omega_H\neq 0$) solutions (see Fig.\ref{JWSJ}).
Since by changing $\Omega_H \rightarrow -\Omega_H$ the energy does not change, for fixed values of the charge there are two  such $J=0$ solutions.

In Fig. \ref{EJ;alpha,Q}, as $\alpha=0$ for both negative and positive values of charge we get the same curve that is no longer valid for non-zero CS coupling constant. We see that for large values of $J$ the $\alpha=0$ curve eventually tends to the KAdS curve.

%%%%%%%%%%%%%%%%%%%%%%%%%%%%%%%%%%%%%%%%%%%%%%%%%%%%%%%%%%%%%%%
%\graphicspath{ {E:/mozhganpro1/newpaper/plotq3eps/} }
\begin{figure*}[h]
\centering
\begin{tabular}{cc}
\includegraphics[scale=.2]{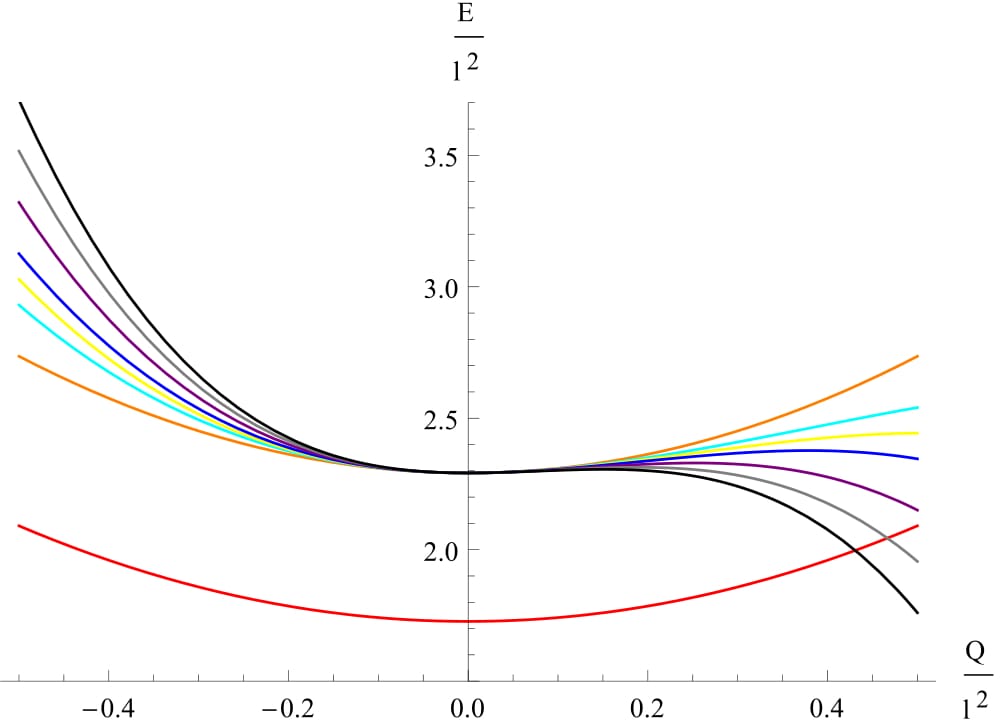}&
\includegraphics[scale=.2]{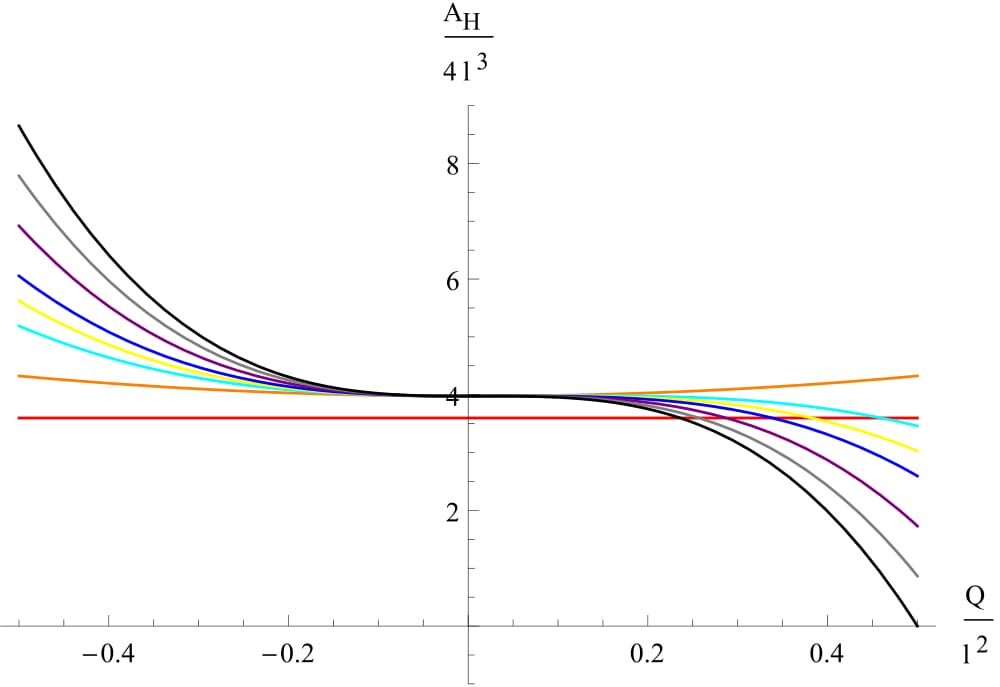}\\
\includegraphics[scale=.2]{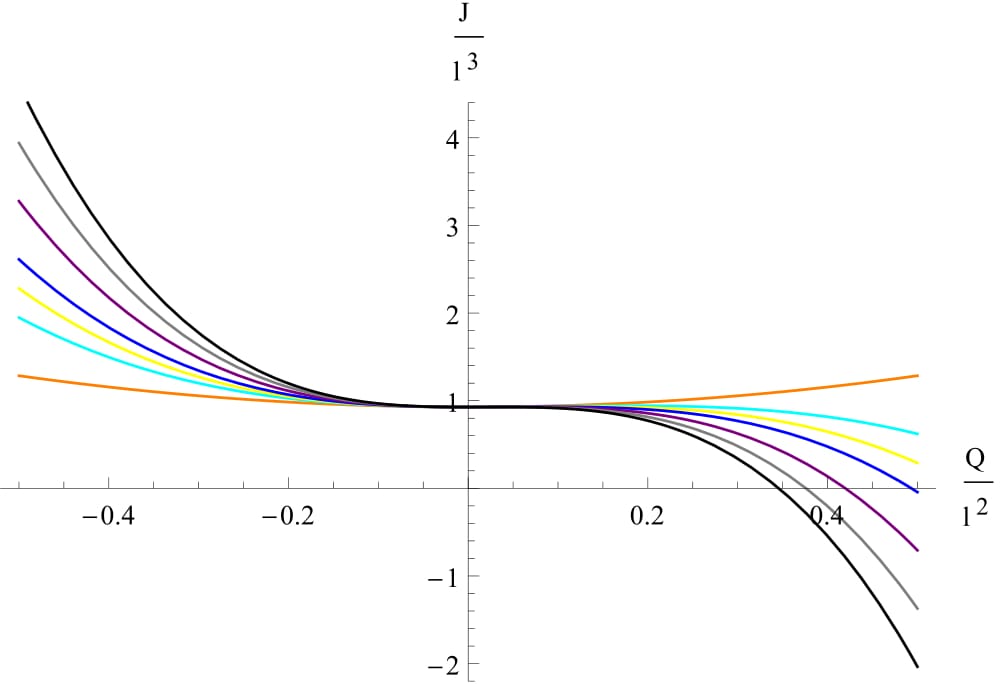}&
\includegraphics[scale=.2]{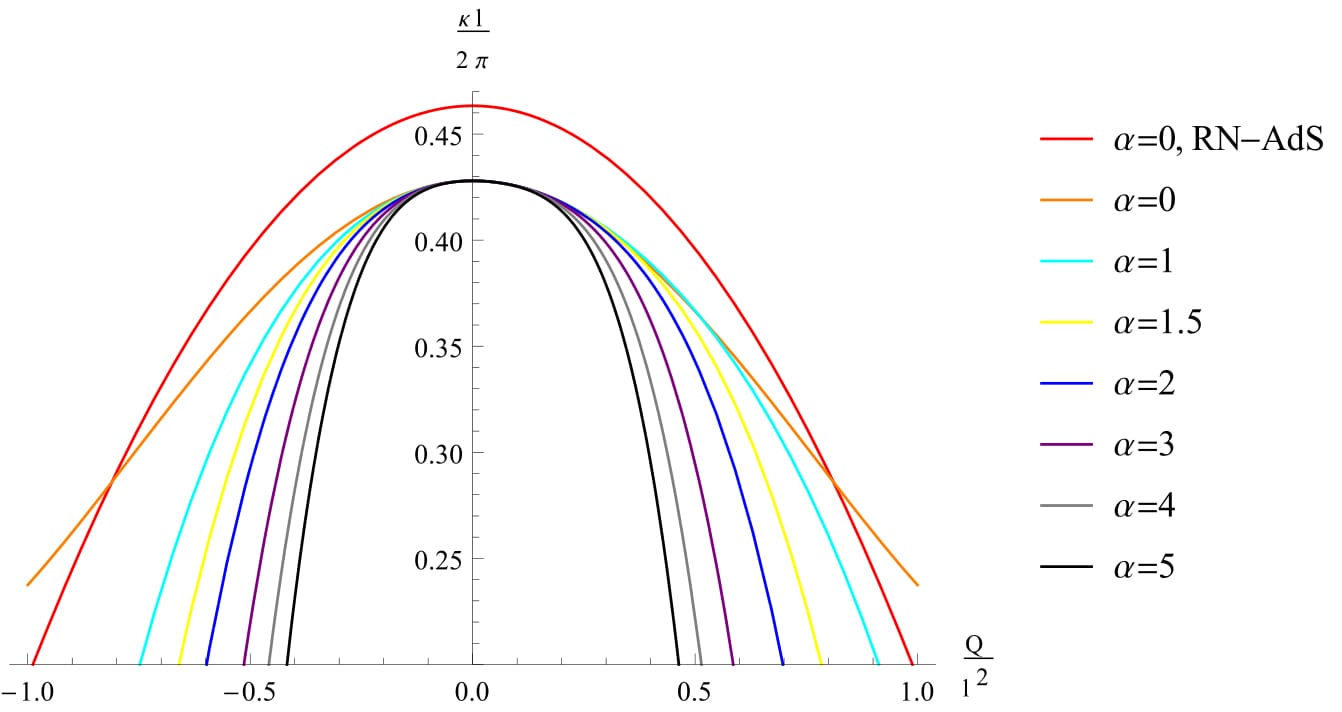}
\\
\end{tabular}
\caption{Energy (upper left),   entropy (upper right),  angular momentum (lower left), and
 temperature (lower right) plotted as a function of charge, with the $J=0$ RN-AdS results plotted
 for comparison, with $r_+/l=0.9 $ and $\Omega_H l = 0.717879$.
In the upper-left figure the intersection point at $Q=0$ is at $E/l^2 = 2.29188$ corresponding to the
KAdS solution for all values of $\alpha$, whereas the corresponding point for the RN-AdS solution is
at $E/l^2 = 1.72721$.   In the remaining figures the $Q=0$ intersection points are at
$A_H/4l^3 = 3.97521$ (with the RN-AdS intersection at $A_H/4l^3 =3.59747$)
 $J/l^3 = 0.927443$, and $\kappa l/2\pi =  0.360645$ (with the RN-AdS intersection at $\kappa l/2\pi=0.463318$).
}\label{EAJT;Q}
\end{figure*}

Next in figure \ref{EAJT;Q} we plot energy, area (or entropy), angular momentum, and temperature
(surface gravity) as a function of charge, the latter including $Q^4/l^8$ terms.  We see that for $Q<0$ all quantities  monotonically increasing functions of $|Q|$ except for the surface gravity, which monotonically decreases.  However
for $Q>0$ the situation is markedly different, with $E$, $J$   increasing with increasing
$Q$ for small $\alpha$ but decreasing with increasing $Q$ for large $\alpha > 1$. The counterrotation
phenomenon is again evident with $J<0$ above some threshold value of $Q$ for sufficiently large $\alpha$.
Likewise   there exists a maximal value of $Q$ at which the area vanishes (fig.  \ref{EAJT;Q}(b)) for large enough $\alpha>2$. Note that
since we chose  $\Omega_H l=0.717879$ for all graphs in order to make a comparison, in the given range of charge on the solution with the largest value of $\alpha$ ultimately attains zero horizon area.% The same situation holds for the plot pertaining to the temperature ($\kappa$).

In Fig. \ref{EAJT;Q}(d) (lower-right),  we see for all values of $\alpha$ that $T$ is a decreasing function of $Q$ that reaches zero.
 Extremal black holes form the lower part of the boundary, since they have $\kappa=0$. The uppermost curve is the RN-AdS solution, included for comparison.

\begin{figure*}[h]
\centering
\begin{tabular}{cc}
\includegraphics[scale=.17]{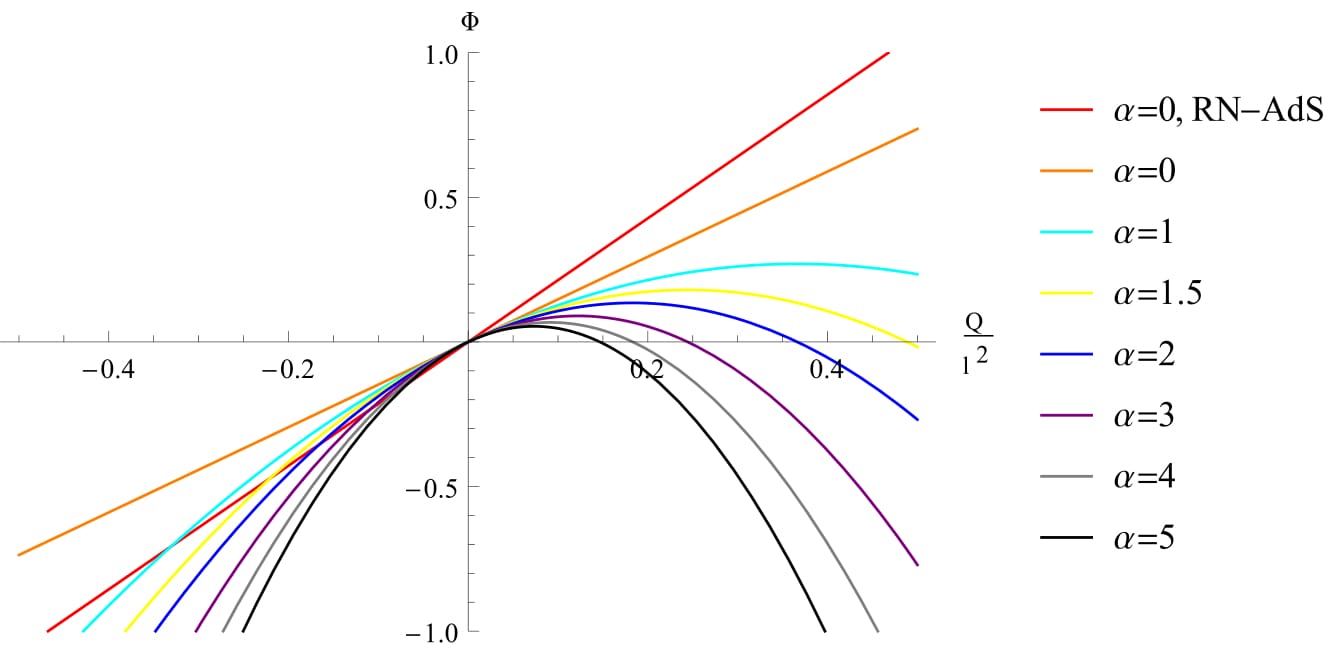}&
\includegraphics[scale=.17]{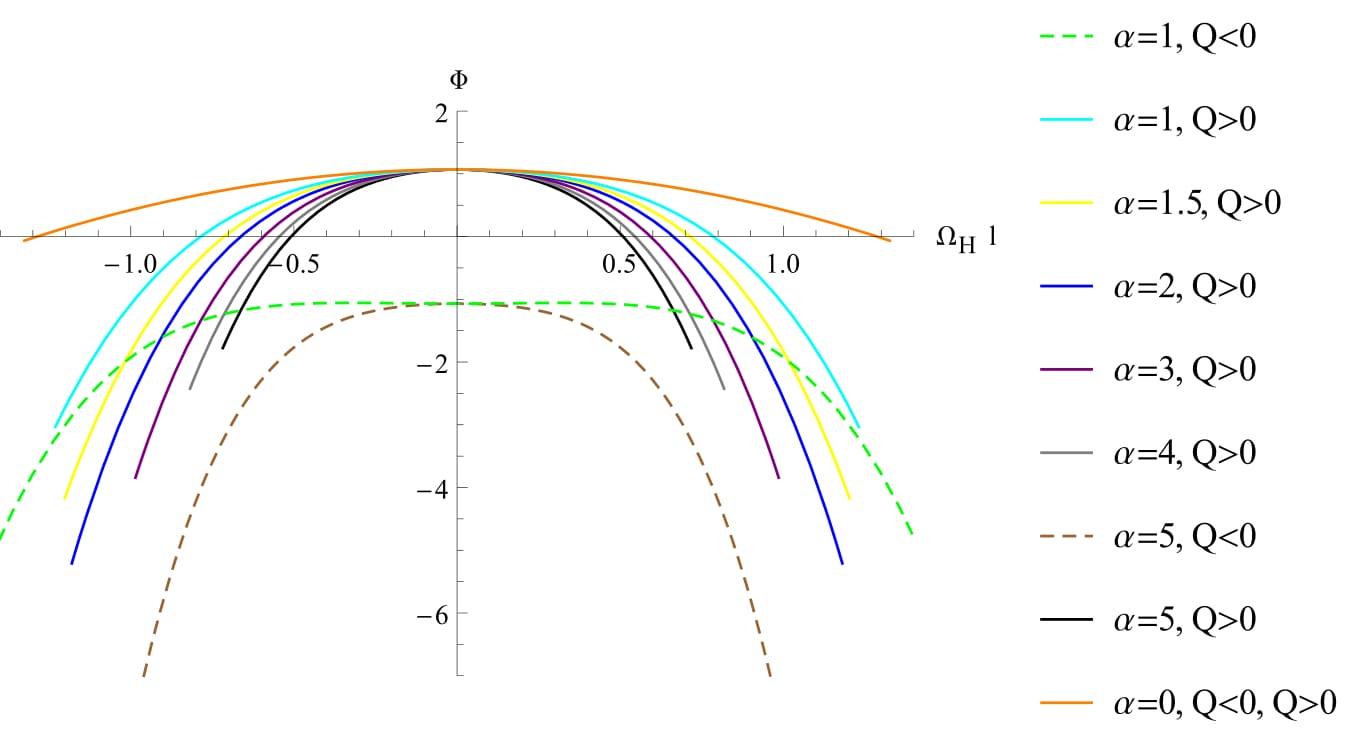} \\
\end{tabular}
\caption{ The electrostatic potential versus the charge (left) for fixed $\Omega_H l = 0.717879$
and the horizon angular velocity (right) for fixed $|Q|/\ell^2 = 0.5$. Curves for positive and negative charges are presented by solid and dashed lines respectively.}
\label{ch;QW}
\end{figure*}
In figure \ref{ch;QW} we see for $Q>0$ and $\alpha<1$ the electrostatic potential is an initially increasing function of $Q$; for larger values of $\alpha$ it begins to decrease eventually changing  sign. For $Q<0$,  $|\Phi|$ increases with increasing $|Q|$.  The right-hand part of shows that
the electrostatic potential  is a decreasing function of  $\Omega_H$.

\begin{figure*}[h]
\centering
\begin{tabular}{cc}
\includegraphics[scale=.19]{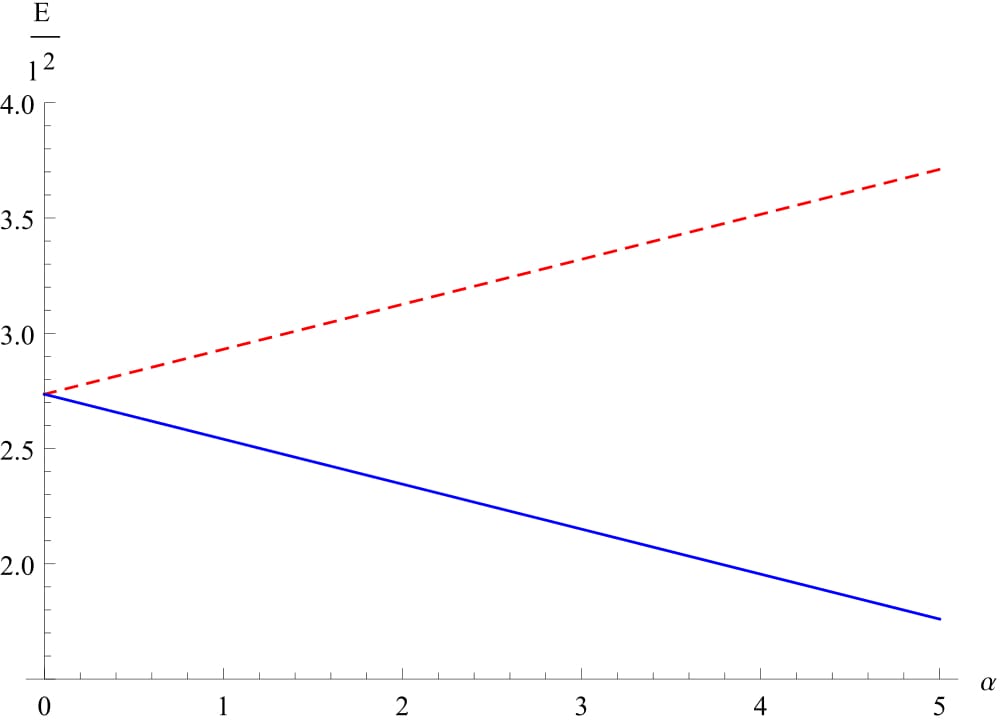}& \quad\quad
\includegraphics[scale=.19]{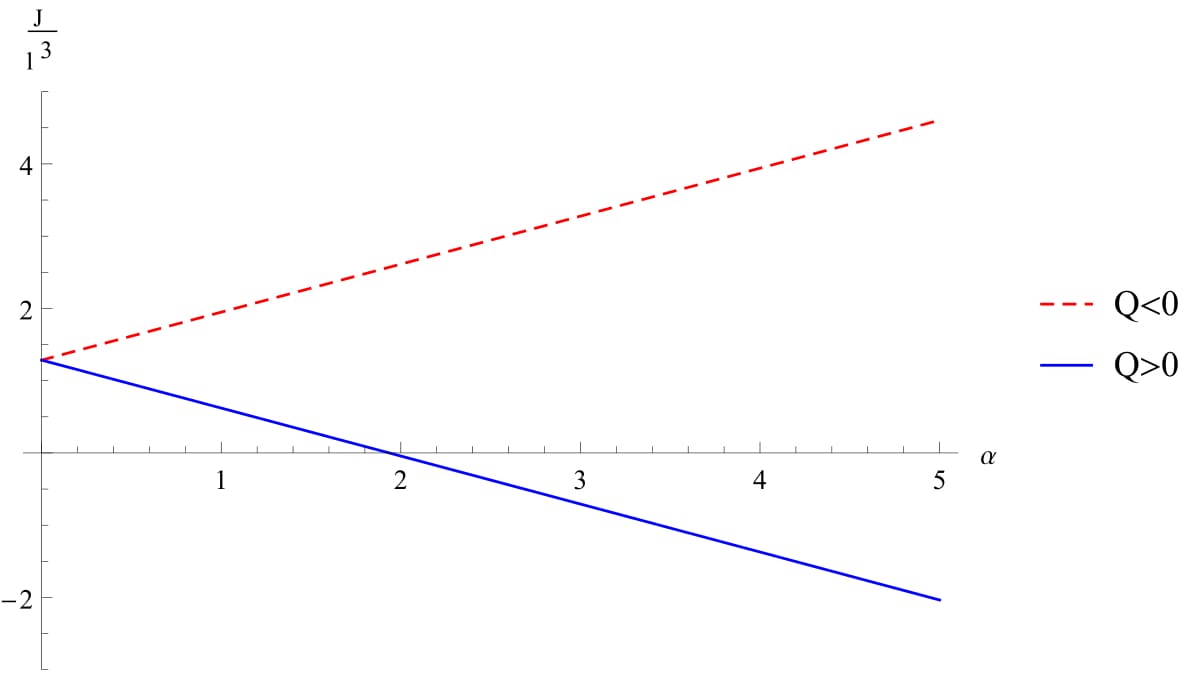} \\
\end{tabular}
\caption{The energy (left) and the entropy (right) versus the CS coupling
for $|Q|/l^2= 0.5$, $r_+/l=0.9$, $\Omega_H l=0.717879 $.
In Fig. (a) the intersection of two lines is given by $E/l^2=2.73574$. In Fig. (b) The joining point is given by $J/l^3=1.28512$.
Curves for positive and negative charges are presented by solid and dashed lines respectively.}\label{EJ;AQ}
\end{figure*}

 Finally in fig. \ref{EJ;AQ} we illustrate the dependence of $E$ and $J$ on $\alpha$.
 In Fig. \ref{EJ;AQ} (a) for positive charge, with increasing $\alpha$ , $E$ decreases whereas for
negative charge it grows. For $\Omega_H l=0.717879$ turns out that as $\alpha > 1.93561$, $J$ becomes negative, indicative of counterrotation for $Q>0$ that is absent for $Q<0$.

\section{Conclusion} \label{conclusion}

We have obtained perturbative charged rotating AdS black hole solutions for Einstein-Maxwell-Chern-Simons theory for arbitrary values of the coupling $\alpha = \lambda/\lambda_{SG}$. The choice of equal magnitude angular momenta
factorizes the angular dependence in the equations, yielding  ordinary differential equations that are difficult but tractable.
By perturbing  about the Kerr-AdS solutions in 5 dimensions with electric charge $Q$ and horizon radius $r_+$ as expansion parameters, we obtained series solutions in both the near and far regions. Our fourth order results are in agreement with the known supergravity solution presented in the Appendix \ref{pertpope} as $\alpha=1$, verifiing the validity of our perturbative approach.  Our results
also yield charged doubly-equally-rotating AdS black holes in Einstein-Maxwell theory by setting  $\alpha=0$.

Using these expansions we extract physical properties of the black holes.
We find that interesting features  emerge at cubic order in the charge, and observe that the behaviour of physical quantities depends strongly on the value of $\alpha$ and the sign of   $Q$,  qualitatively in agreement with  numerical solutions previously obtained \cite{Kunz:2005,Kunz:2006B}. Although $Q<0$ solutions do not reveal any peculiar behaviour, the positive charge solutions present intriguing properties once the coupling constant  exceeds its supergravity value ($\alpha > 1$).

Scaling all quantities in units of the AdS length,
we observe the phenomenon of counter-rotation  \cite{Kleihaus:2004}, with the signs of $J$ and $\Omega_H$ opposite for sufficiently large $\alpha$.
 For $\alpha>1$, non-static $J=0$ black holes exist, and upon setting $J\rightarrow -J$ we find two such stationary solutions. They have the same area horizon angular velocities of but opposite sign. Other physical charges remain the same, indicative of non-uniqueness \cite{Blazquez-Salcedo:2013muz}.
Upon increasing $\alpha$ beyond its supergravity value, the horizon area starts to decrease with increasing $J$ and eventually vanishes, signalling the onset of an  instability \cite{Myers:1999,Herdeiro:2000}. For large enough $\alpha$, it goes to zero and the rotating $J=0$ solutions disappear.

 Although we worked in 5 dimensions, our procedure is applicable to arbitrary
odd dimensional solutions in EMCS theory, when all the angular momenta are equal.
 We anticipate similar properties in a such theory upon including scalars.

\section*{Acknowledgements}
MM would like to thank  Francisco Navarro-L´erida for very useful comments and M.M. Sheikh-Jabbari for discussions; RBM would like to thank Jose Blasquez-Salcedo for interesting discussions as well. MM appreciates the hospitality of University of Waterloo during accomplishing this work.
This work was supported in part by the Natural Sciences and Engineering Council of Canada.

\section*{Note added}
During the preparation of this paper for publication, we received communication from J. Blasquez-Salcedo, J.Kunz,  F. Navarro-Lerida, and E. Radu concerning their work \cite{Jose}
which overlaps with  some results that we have obtained.  Their approach is complementary to ours:
they obtain numerical solutions to the EMCS equations whose asymptotic structure agrees with our solutions.

\appendix

\section{ Field Equations} \label{odeq}

Here we present the field equations and the gauge equations \eqref{Eineq}, simplified using
our ansatz.  The quantities $C_1$ and $C_2$ are the constraints that must also be satisfied.  Our perturbative solutions to these equations appear in subsequent appendices.
\beqa
C_1 &\equiv&g(r) \left(h(r) \left(2 r f'(r)+4 f(r)-\frac{8 r^2}{l^2}-8\right)+r f(r) h'(r)+4 h(r)^2\right)+r f(r) h(r) g'(r)
\nonumber\\
&&\left.+\frac{1}{3 r^2}\left(r^4 h(r) \left(a_0'(r)+\Omega (r) a_1'(r)\right)^2-r^2 f(r) g(r) a_1'(r)^2+8 a_1(r)^2 g(r) h(r)\right)=0\right.\nonumber
\\
C_2 &\equiv&\frac{1}{f(r)}\left(\frac{3 f'(r)}{2 r}-\frac{r^2 h(r) \Omega '(r)^2}{4 g(r)}+\frac{h'(r)}{r}+\frac{5 h(r)}{r^2}-\frac{6}{l^2}-\frac{8}{r^2}\right)+\frac{h'(r)}{f(r) h(r)} \Big(\frac{f'(r)}{4}-\frac{2 r}{l^2}\nonumber\\
&&\left.-\frac{2}{r}\Big)
+\frac{h'(r)}{h(r)} \left(\frac{h'(r)}{4 h(r)}+\frac{3}{2 r}\right)+\frac{3}{r^2}+\frac{1}{12 r^4 f(r) g(r) h(r)^2}\Big(r^2 \left(r h'(r)+3 h(r)\right)
\Big(r^2 h(r)\right.\nonumber\\
&&\left.\times \left(a_0'(r)+\Omega (r) a_1'(r)\right)^2-f(r) g(r) a_1'(r)^2\Big)+4 a_1(r)^2 g(r) h(r) \left(2 r h'(r)+9 h(r)\right)\Big)=0\right.\nonumber
\\
&&h''(r)+h'(r)\left(\frac{1}{f(r)}\left(-\frac{2 h(r)}{r}+\frac{4 r}{l^2}+\frac{4}{r}\right)+\frac{1}{r}\right)-\frac{h'(r)^2}{h(r)}+\frac{h(r) }{f(r)} \Big(\frac{r^2 h(r) \Omega '(r)^2}{g(r)}\nonumber\\
&&\left.-\frac{8 h(r)}{r^2}+\frac{8}{r^2}\Big)-\frac{1}{6 r^4 f(r) g(r) h(r)}\Big(r^2 \Big(r^3 h(r) h'(r) \left(a_0'(r)+\Omega (r) a_1'(r)\right)^2-f(r) g(r)\right.\nonumber\\
&&\left.\times a_1'(r)^2 \big(r h'(r)+6 h(r)\big)\Big)
+8 a_1(r)^2 g(r) h(r) \left(r h'(r)+3 h(r)\right)\Big)=0\right.\nonumber
\eeqa
\beqa
&&\Omega ''(r)+\Omega '(r)\left(\frac{1}{f(r)}\left(f'(r)+\frac{2 h(r)}{r}-\frac{4 r}{l^2}-\frac{4}{r}\right)+\frac{2 h'(r)}{h(r)}+\frac{7}{r}\right)+\frac{1}{6 r^3 f(r) g(r) h(r)}\nonumber\\
&&\left.\times\Big(h(r) \Omega '(r) \left(r^4 \left(a_0'(r)+\Omega (r) a_1'(r)\right)^2+8 a_1(r)^2 g(r)\right)-r f(r) g(r) a_1'(r) \big(6 a_0'(r)\right.\nonumber\\
&&\left.+a_1'(r) \big(r \Omega '(r)+6 \Omega (r)\big)\big)\Big)=0\right. \nonumber
\\
&&f''(r)-\frac{f'(r)^2}{f(r)}+f'(r)\left(\frac{1}{f(r)}\left(-\frac{6 h(r)}{r}+\frac{12 r}{l^2}+\frac{12}{r}\right)+\frac{1}{r}\right)+\frac{1}{f(r)}\Big(8 h(r) \Big(-\frac{h(r)}{r^2}\nonumber\\
&&\left.+\frac{4}{l^2}+\frac{4}{r^2}\Big)
-32 \left(\frac{r^2}{l^4}+\frac{2}{l^2}+\frac{1}{r^2}\right)\Big)+\frac{4 h'(r)}{r}+\frac{8}{l^2}+\frac{8}{r^2}+\frac{1}{6 l^2 r^4 f(r) g(r)^2 h(r)^2}\bigg(r^2\right.\nonumber\\
&&\left.\times \big(-l^2 r^3 f(r) h(r)^2 g'(r)  \left(a_0'(r)+\Omega (r) a_1'(r)\right)^2+r g(r) h(r) \bigg(l^2 f(r) \Big(r^2 h'(r) \big(a_0'(r)\right.\nonumber\\
&&\left.+\Omega (r) a_1'(r)\big)^2-f(r) a_1'(r)^2 g'(r)\Big)-4 l^2 r h(r)^2 \left(a_0'(r)+\Omega (r) a_1'(r)\right)^2+r h(r) \big(a_0'(r)\right.\nonumber\\
&&\left.+\Omega (r) a_1'(r)\big) \Big(4 l^2 r  f(r) \left(a_0''(r)+a_1'(r) \Omega '(r)\right)+a_0'(r) \left(8 \left(l^2 f(r)+l^2+r^2\right)-l^2 r f'(r)\right)\right.\nonumber\\
&&\left. +\Omega (r) \Big(4 l^2 r f(r) a_1''(r)+a_1'(r) \big(8 \left(l^2 f(r)+l^2+r^2\right)-l^2 r f'(r)\big)\Big)\Big)\Big)+f(r) g(r)^2 a_1'(r) \right.\nonumber\\
&&\left.\times \Big(a_1'(r) \Big(h(r) \left(l^2 \left(-\left(r f'(r)+4 f(r)-4 h(r)\right)\right)-8 \left(l^2+r^2\right)\right)
+l^2 r f(r) h'(r)\Big)\right.\nonumber
\\
&&\left.-4 l^2 r f(r) h(r) a_1''(r)\Big)\bigg)+32 l^2 r a_1(r)  f(r) g(r)^2 h(r)^2 a_1'(r)+8 a_1(r)^2 g(r) h(r) \Big(g(r) \Big(h(r)\right.\nonumber\\
&&\left.\times \big(8 \left(l^2+r^2\right)-l^2 \big(r f'(r)+4 h(r)\big)\big)+l^2 r f(r) h'(r)\Big)+l^2 r f(r) h(r) g'(r)\Big)\bigg)=0\right.\nonumber
\\
&&g''(r)-\frac{g'(r)^2}{g(r)}+g'(r) \left(\frac{1}{f(r)}\left(-\frac{4 h(r)}{r}+\frac{8 r}{l^2}+\frac{8}{r}\right)+\frac{1}{r}\right)
+\frac{1}{f(r)}\Big(g(r) \Big(h(r) \Big(-\frac{r^2 \Omega '(r)^2}{g(r)}\nonumber\\
&&\left.-\frac{8}{r^2}\Big)+\left(\frac{4 r}{l^2 h(r)}+\frac{4}{r h(r)}-\frac{6}{r}\right) h'(r)+\frac{8}{r^2}\Big)\Big)-\frac{1}{6 r^4 f(r) h(r)^2}\bigg(r^2 \Big(h(r) \big(r^3 h'(r) \big(a_0'(r)\right.\nonumber\\
&&\left.+\Omega (r) a_1'(r)\big)^2-2 r g(r) a_1'(r)^2 f'(r)-2 f(r) a_1'(r) \left(2 r g(r) a_1''(r)+a_1'(r) \left(r g'(r)+g(r)\right)\right)\big)\right.\nonumber\\
&&\left.+4 r^2 h(r)^2 \left(a_0'(r)+\Omega (r) a_1'(r)\right) \big(r a_0''(r)+3 a_0'(r)+r \Omega (r) a_1''(r)+r a_1'(r) \Omega '(r)+3 \Omega (r) \right.\nonumber\\
&&\left.\times a_1'(r)\big)+r f(r) g(r) a_1'(r)^2 h'(r)\Big)+32 r a_1(r) g(r) h(r)^2 a_1'(r)+8 a_1(r)^2 h(r) \big(2 r h(r) g'(r)\right.\nonumber\\
&&\left.+g(r) \big(r h'(r)+h(r)\big)\big)\bigg)=0\right.\nonumber
\\
&&a_0''(r)+a_1'(r) \Big(a_1(r) \sqrt{g(r) h(r)} \left(\frac{8 \lambda }{r^3 h(r)}-\frac{8 \lambda  \Omega (r)^2}{r f(r) g(r)}\right)-\frac{\Omega (r) f'(r)}{f(r)}+\frac{r^2 h(r) \Omega (r)^2 \Omega '(r)}{f(r) g(r)}\nonumber\\
&&\left.-\frac{\Omega (r) g'(r)}{g(r)}+\frac{\Omega (r) h'(r)}{h(r)}+\Omega '(r)+\frac{2 \Omega (r)}{r}\Big)+\frac{4 a_1(r) h(r) \Omega (r)}{r^2 f(r)}+a_0'(r) \Big(-\frac{8 \lambda  a_1(r) \Omega (r)}{r f(r) g(r)} \right.\nonumber\\
&&\left.\times\sqrt{g(r) h(r)}+\frac{r^2 h(r) \Omega (r) \Omega '(r)}{f(r) g(r)}-\frac{g'(r)}{2 g(r)}+\frac{h'(r)}{2 h(r)}+\frac{3}{r}\Big)=0\right.\nonumber
\eeqa
\beqa
&&a_1''(r)+\frac{h(r)}{r^2 f(r)} \left(a_1(r) \left(\frac{8 \lambda  r a_0'(r)}{\sqrt{g(r) h(r)}}-4\right)-\frac{r^4 a_0'(r) \Omega '(r)}{g(r)}\right)+a_1'(r)\Big(\frac{8 \lambda  a_1(r) \Omega (r)}{r f(r) g(r)}\nonumber\\
&&\left.\times \sqrt{g(r) h(r)}+\frac{f'(r)}{f(r)}-\frac{r^2 h(r) \Omega (r) \Omega '(r)}{f(r) g(r)}+\frac{g'(r)}{2 g(r)}-\frac{h'(r)}{2 h(r)}+\frac{1}{r}\Big)=0\right.\nonumber
\eeqa

\section{Perturbative results} \label{pert}

Here we present our results for solving the field equations \eqref{Eineq} perturbatively to order $Q^4/l^8$,
retaining relevant powers of $r_+/l$ as appropriate.
In the following, $\text{Li}_n(z)$ denotes the polylog function,  defined as
\beqa
\text{Li}_n(z)=\sum _{k=1}^{\infty } \frac{z^k}{k^n}
\eeqa
\subsection{Far-field Expansion}

In the far-field expansion we obtain
%Form the expansion of KAdSAdS solution at order $Q=0$ we get
\beqa
f^{out}(r)&=&1+\frac{r^2}{l^2}-\frac{l^2}{r^2}\left(\frac{r_+^2}{l^2}\right)-\frac{l^2 \left(l^2 \Omega_H^2+1\right)}{r^2} \left(\frac{r_+^4}{l^4}\right)+\frac{l^4 \Omega_H^2}{r^4} \left(l^2 \left(1-\Omega_H^2 r^2\right)+r^2\right) \left(\frac{r_+^6}{l^6}\right)\nonumber\\
&&\left.-\frac{l^4 \Omega_H^2}{r^4} \left(l^2 \Omega_H^2-1\right) \left(l^2 \left(\Omega_H^2 r^2-1\right)-r^2\right) \left(\frac{r_+^8}{l^8}\right)+\cO\left(\frac{r_+^{10}}{l^{10}}\right)+\left(\frac{Q^2}{l^4}\right)
\bigg(-\frac{l^2}{r^2}\right.\nonumber\\
&&\left.\times\left(\frac{l^{2}}{r_+^{2}}\right)+\frac{l^4}{r^4}+\frac{l^6 \Omega_H^4}{r^2}\left(\frac{r_+^2}{l^2}\right)+\frac{l^4 \Omega_H^2}{r^4} \left(r^2 \left(l^2 \Omega_H^2 \left(l^2 \Omega_H^2-4\right)+1\right)-l^4 \Omega_H^2\right)\left(\frac{r_+^4}{l^4}\right)\right.\nonumber\\
&&\left.+\cO\left(\frac{r_+^{6}}{l^{6}}\right)\bigg)+\left(\frac{Q^3}{l^6}\right)\bigg(-\frac{4 \sqrt{3} \lambda  l^6 \Omega_H^4}{r^2}+\frac{4 \sqrt{3} \lambda  l^4 \Omega_H^2}{r^4} \left(l^4 \Omega_H^2+r^2 \left(5 l^2 \Omega_H^2-1\right)\right)\right.\nonumber\\
&&\left.\times\left(\frac{r_+^2}{l^2}\right)+\cO\left(\frac{r_+^{4}}{l^{4}}\right)\bigg)+\left(\frac{Q^4}{l^8}\right)\left(\frac{2 \left(12 \lambda ^2-1\right) l^4 \Omega_H^2}{9 r^2} \left(\frac{l^{4}}{r_+^{4}}\right)+\cO\left(\frac{l^{2}}{r_+^{2}}\right)\right)\right.\nonumber
\\
g^{out}(r)&=&1-\frac{l^6 \Omega_H^2}{r^4}\left(\frac{r_+^6}{l^6}\right)-\frac{l^6 \Omega_H^2 \left(l^2 \Omega_H^2-1\right)}{r^4}\left(\frac{r_+^8}{l^8}\right)
+\cO\left(\frac{r_+^{10}}{l^{10}}\right)+\left(\frac{Q^2}{l^4}\right)
\bigg(\frac{l^6 \Omega_H^2}{r^6} \big(r^2\nonumber\\
&&\left.\times \left(l^2 \Omega_H^2-1\right)+l^2\big)\left(\frac{r_+^4}{l^4}\right)+\cO\left(\frac{r_+^{6}}{l^{6}}\right)\bigg) -\left(\frac{Q^3}{l^6}\right)\bigg(\frac{4 \sqrt{3} \lambda  l^6 \Omega_H^2}{r^6} \big(r^2 \left(l^2 \Omega_H^2-1\right)\right.\nonumber\\
&&\left.+l^2\big)\left(\frac{r_+^2}{l^2}\right)+\cO\left(\frac{r_+^{4}}{l^{4}}\right)\bigg)
+\left(\frac{Q^4}{l^8}\right)\left(\cO\left(\frac{l^{2}}{r_+^{2}}\right)\right)
\right.\nonumber
\\
h^{out}(r)&=&1+\frac{l^6 \Omega_H^2}{r^4}\left(\frac{r_+^6}{l^6}\right)+\frac{l^6 \Omega_H^2}{r^4} \left(l^2 \Omega_H^2-1\right)\left(\frac{r_+^8}{l^8}\right)
+\cO\left(\frac{r_+^{10}}{l^{10}}\right)-\left(\frac{Q^2}{l^4}\right)
\bigg(\frac{l^6 \Omega_H^2}{r^6} \big(r^2 \nonumber
\eeqa
\beqa
&&\left.\times\left(l^2 \Omega_H^2-1\right)+l^2\big)\left(\frac{r_+^4}{l^4}\right)+\cO\left(\frac{r_+^{6}}{l^{6}}\right)\bigg)+\left(\frac{Q^3}{l^6}\right)
\bigg(\frac{4 \sqrt{3} \lambda  l^6 \Omega_H^2}{r^6} \big(r^2 \left(l^2 \Omega_H^2-1\right)\right.\nonumber\\
&&\left.+l^2\big)\left(\frac{r_+^2}{l^2}\right)+\cO\left(\frac{r_+^{4}}{l^{4}}\right)\bigg)
+\left(\frac{Q^4}{l^8}\right)\left(\cO\left(\frac{l^{2}}{r_+^{2}}\right)\right)
\right.\nonumber
\\
\Omega^{out}(r)&=&\frac{l^4 \Omega_H}{r^4}\left(\frac{r_+^4}{l^4}\right)+\frac{l^6 \Omega_H^3}{r^4}\left(\frac{r_+^6}{l^6}\right)+\frac{l^6 \Omega_H^3 \left(l^2 \Omega_H^2-1\right)}{r^4}\left(\frac{r_+^8}{l^8}\right)+\left(\frac{Q^2}{l^4}\right)\bigg(\frac{l^4 \Omega_H}{r^4}-\frac{l^4 \Omega_H}{r^6}\nonumber\\
&&\left.\times \left(l^2 \left(\Omega_H^2 r^2+1\right)+r^2\right)\left(\frac{r_+^2}{l^2}\right)+\frac{l^4 \Omega_H}{r^6} \left(r^2 \left(1-l^2 \Omega_H^2 \left(l^2 \Omega_H^2-3\right)\right)+l^2\right)\left(\frac{r_+^4}{l^4}\right)\right.\nonumber\\
&&\left.+\cO\left(\frac{r_+^{6}}{l^{6}}\right)\bigg) +\left(\frac{Q^3}{l^6}\right)\bigg(-\frac{2 \sqrt{3} \lambda  l^4 \Omega_H}{r^4}\left(\frac{l^2}{r_+^2}\right)+\frac{2 \sqrt{3} \lambda  l^4 \Omega_H}{r^6} \big(r^2 \left(3 l^2 \Omega_H^2+2\right)\right.\nonumber\\
&&\left.+l^2\big)-\frac{2 \sqrt{3} \lambda  l^4 \Omega_H }{r^6}\left(l^2 \left(\Omega_H^2 \left(l^2+11 r^2\right)+2\right)+3 r^2\right)\left(\frac{r_+^2}{l^2}\right)+\cO\left(\frac{r_+^{4}}{l^{4}}\right)\bigg) \right.\nonumber\\
&&\left.+\left(\frac{Q^4}{l^8}\right)\left(\frac{12 \lambda ^2 l^4 \Omega_H}{r^4}\left(\frac{l^4}{r_+^4}\right)+\cO\left(\frac{l^2}{r_+^2}\right)\right)\right.\nonumber
\eeqa
for the metric functions, and
\beqa
a^{out}_{0}(r)&=&\left(\frac{Q}{l^2}\right)\left(\frac{\sqrt{3} l^2 }{r^2}+\cO\left(\frac{r_+^{10}}{l^{10}}\right)\right)+
\left(\frac{Q^2}{l^4}\right)\left(\cO\left(\frac{r_+^{6}}{l^{6}}\right)\right)+
\left(\frac{Q^3}{l^6}\right)\left(\cO\left(\frac{r_+^{2}}{l^{2}}\right)\right)\nonumber\\
&&\left.+
\left(\frac{Q^4}{l^8}\right)\left(\cO\left(\frac{l^{2}}{r_+^{2}}\right)\right)\right.
\label{a0out}
\\
a^{out}_{1}(r)&=&\left(\frac{Q}{l^2}\right)\left(\frac{\sqrt{3} l^4 \Omega_H}{r^2}\left(-\left(\frac{r_+^2}{l^2}\right)+\left(\frac{r_+^4}{l^4}\right)-\left(\frac{r_+^6}{l^6}\right)+\left(\frac{r_+^8}{l^8}\right)\right) +\cO\left(\frac{r_+^{10}}{l^{10}}\right)\right) \nonumber\\
&&\left.+\left(\frac{Q^2}{l^4}\right)\bigg(\frac{6 \lambda  l^4 \Omega_H}{r^2}-\frac{6 \lambda  l^4 \Omega_H}{r^2} \left(l^2 \Omega_H^2+2\right)\left(\frac{r_+^2}{l^2}\right)+\frac{18 \lambda  l^4 \Omega_H}{r^2} \left(l^2 \Omega_H^2+1\right)\left(\frac{r_+^4}{l^4}\right)\right.\nonumber\\
&&\left.+\cO\left(\frac{r_+^{6}}{l^{6}}\right)\bigg) +\left(\frac{Q^3}{l^6}\right)\bigg(-\frac{12 \sqrt{3} \lambda ^2 l^4 \Omega_H}{r^2}\left(\frac{l^{2}}{r_+^{2}}\right)+\frac{l^4 \Omega_H}{\sqrt{3} r^2} \big(108 \lambda ^2+\left(96 \lambda ^2+1\right)\right.\nonumber\\
&&\left.\times l^2 \Omega_H^2\big)+\cO\left(\frac{r_+^{2}}{l^{2}}\right)\bigg)+\left(\frac{Q^4}{l^8}\right)\Big(\frac{72 \lambda ^3 l^4 \Omega_H}{r^2} \left(\frac{l^4}{r_+^4}\right)+\cO\left(\frac{l^{2}}{r_+^{2}}\right)\Big)\right.\nonumber
\eeqa
for the gauge field.  It is straightforward to check that we recover the results in section \ref{FFP} upon setting
$\lambda = \lambda_{SG}$.

\subsection{Near-field Expansion}

As for the exact case, the near-field expansions are considerably lengthier, and so we present the successive terms in
the expansion for each function.

For the metric functions we find that
\beqa
f^{in}_0(z) &=&1-\frac{l^2}{z^2}+\frac{1}{l^2 z^4}\left(l^8 \Omega_H^2-l^4 z^2 \left(l^2 \Omega_H^2+1\right)+z^6\right)\left(\frac{r_+^2}{l^2}\right)+\frac{l^4 \Omega_H^2}{z^4} \left(l^2 \Omega_H^2-1\right) \left(l^2-z^2\right)\nonumber\\
&&\left.\times\left(\frac{r_+^4}{l^4}\right) +\frac{l^4 \Omega_H^2}{z^4} \left(l^2 \Omega_H^2-1\right)^2 \left(l^2-z^2\right) \left(\frac{r_+^6}{l^6}\right) +\frac{l^4 \Omega_H^2}{z^4} \left(l^2 \Omega_H^2-1\right)^3 \left(l^2-z^2\right)\right.\nonumber\\
&&\left.\times\left(\frac{r_+^8}{l^8}\right) + \cO\left(\frac{r_+^{10}}{l^{10}}\right)\right.\nonumber
\\
f^{in}_2(z) &=&\left(\frac{Q^2}{l^4}\right)\bigg(\frac{l^2 \left(l^2-z^2\right)}{z^4}\left(\frac{l^4}{r_+^4}\right)-\frac{l^6 \Omega_H^4 \left(l^2-z^2\right)}{z^4}-\frac{l^4 \Omega_H^2}{z^4} \left(l^2 \Omega_H^2 \left(l^2 \Omega_H^2-4\right)+1\right) \left(l^2-z^2\right)\nonumber\\
&&\left.\times\left(\frac{r_+^2}{l^2}\right)-\frac{l^4 \Omega_H^2}{z^4} \left(l^2 \Omega_H^2 \left(l^2 \Omega_H^2 \left(l^2 \Omega_H^2-5\right)+10\right)-3\right) \left(l^2-z^2\right)\left(\frac{r_+^4}{l^4}\right)+\cO\left(\frac{r_+^{6}}{l^{6}}\right)\bigg)\right.\nonumber
\\
f^{in}_3(z) &=&\left(\frac{Q^3}{l^6}\right)\bigg(\frac{4 \sqrt{3} \lambda  l^6 \Omega_H^4}{z^4} \left(l^2-z^2\right)\left(\frac{l^2}{r_+^2}\right)+\frac{4 \sqrt{3} \lambda  l^4 \Omega_H^2}{z^4} \left(1-5 l^2 \Omega_H^2\right) \left(l^2-z^2\right)\nonumber\\
&&\left.+\frac{4 \sqrt{3} \lambda  l^4 \Omega_H^2}{z^4}\left(14 l^2 \Omega_H^2-4\right) \left(l^2-z^2\right)\left(\frac{r_+^2}{l^2}\right) +\cO\left(\frac{r_+^{4}}{l^{4}}\right)\bigg)\right.\nonumber
\\
f^{in}_4(z) &=&\left(\frac{Q^4}{l^8}\right)\bigg(-\frac{\left(12 \lambda ^2-1\right) l^4 \Omega_H^2}{9 z^8} \left(5 l^6-6 l^4 z^2+3 l^2 z^4-2 z^6\right)\left(\frac{l^6}{r_+^6}\right)+\frac{\Omega_H^2}{18 z^{10}}\Big(l^2 \big(6 \nonumber\\
&&\left.\times\left(1-12 \lambda ^2\right) l^{12} \Omega_H^2+11 \left(12 \lambda ^2-1\right) l^{10} \Omega_H^2 z^2-6 \left(12 \lambda ^2-1\right) l^8 z^2 \left(\Omega_H^2 z^2-5\right)\right.\nonumber\\
&&\left.-6 l^6 z^4 \big(96 \lambda ^2
+\left(96 \lambda ^2+1\right) \Omega_H^2 z^2-8\big)+l^4 z^6 \left(432 \lambda ^2+\left(588 \lambda ^2+5\right) \Omega_H^2 z^2-36\right)\right.\nonumber\\
&&\left.-2 \left(81+2 \pi ^2\right) \left(12 \lambda ^2-1\right) l^2 z^8+144 \left(12 \lambda ^2-1\right) z^{10}\big)+24 \left(12 \lambda ^2-1\right) z^8 \Big(\big(2 l^4\right.\nonumber\\
&&\left.-9 l^2 z^2+6 z^4\big) \log \left(1-\frac{l^2}{z^2}\right)-6 l^2 \left(l^2-z^2\right)\Big) \log \left(\frac{z^2}{l^2}\right)-24 \left(12 \lambda ^2-1\right) z^8 \big(2 l^4\right.\nonumber\\
&&\left. -9 l^2 z^2+6 z^4\big) \text{Li}_2\left(\frac{l^2}{z^2}\right)\Big)\left(\frac{l^4}{r_+^4}\right)+\cO\left(\frac{l^{2}}{r_+^{2}}\right)\bigg)\right.\nonumber
\\
g^{in}_0(z) &=&1-\frac{l^6 \Omega_H^2}{z^4}\left(\frac{r_+^2}{l^2}\right) +\frac{l^6}{z^8} \left(l^6 \Omega_H^4-\Omega_H^2 z^4 \left(l^2 \Omega_H^2-1\right)\right)\left(\frac{r_+^4}{l^4}\right) -\frac{l^6}{z^{12}} \big(l^6 \Omega_H^3-\Omega_H z^4 \nonumber\\
&&\left.\times\left(l^2 \Omega_H^2-1\right)\big)^2\left(\frac{r_+^6}{l^6}\right) +\frac{l^6 \Omega_H^2}{z^{16}} \left(l^2 \Omega_H^2 \left(l^4-z^4\right)+z^4\right)^3\left(\frac{r_+^8}{l^8}\right)+\cO\left(\frac{r_+^{10}}{l^{10}}\right)\right.\nonumber
\\
g^{in}_2(z) &=&\left(\frac{Q^2}{l^4}\right)\bigg(\frac{l^8 \Omega_H^2}{z^6}\left(\frac{l^2}{r_+^2}\right)-\frac{l^6 \Omega_H^2}{z^{10}} \left(2 l^2 z^4+l^2 \Omega_H^2 \left(2 l^6-z^6\right)+z^6\right)+\frac{l^6 \Omega_H^2}{z^{14}} \Big(3 l^{14} \Omega_H^4\nonumber\\
&&\left.+z^{10} \left(l^2 \Omega_H^2 \left(l^2 \Omega_H^2-4\right)+3\right)+3 l^2 z^8-2 l^8 \Omega_H^2 z^4
\left(l^2 \Omega_H^2-3\right)-2 l^6 \Omega_H^2 z^6 \big(l^2 \Omega_H^2\right.\nonumber\\
&&\left.-1\big)\Big)\left(\frac{r_+^2}{l^2}\right) +\frac{l^6 \Omega_H^2}{z^{18}} \Big(-4 l^{20} \Omega_H^6+z^{14} \left(l^2 \Omega_H^2 \left(l^2 \Omega_H^2 \left(l^2 \Omega_H^2-5\right)+10\right)-6\right)\right.\nonumber
\\
&&\left.-4 l^2 z^{12}+6 l^{14} \Omega_H^4 z^4 \left(l^2 \Omega_H^2-2\right)+3 l^{12} \Omega_H^4 z^6 \left(l^2 \Omega_H^2-1\right)-2 l^8 \Omega_H^2 z^8 \big(l^2 \Omega_H^2\right.\nonumber
\\
&&\left.\times \left(l^2 \Omega_H^2-4\right)+6\big)-4 l^6 \Omega_H^2 z^{10} \left(l^2 \Omega_H^2 \left(l^2 \Omega_H^2-3\right)+2\right)\Big)\left(\frac{r_+^4}{l^4}\right)+\cO\left(\frac{r_+^{6}}{l^{6}}\right)
\bigg)\right.\nonumber
\eeqa
\beqa
g^{in}_3(z) &=&\left(\frac{Q^3}{l^6}\right)\bigg(-\frac{4 \sqrt{3} \lambda  l^8 \Omega_H^2}{z^6}\left(\frac{l^4}{r_+^4}\right) +\frac{4 \sqrt{3} \lambda  l^6 \Omega_H^2}{z^{10}} \big(z^4 \left(3 l^2+z^2\right)+l^2 \Omega_H^2 \big(2 l^6+l^2 z^4\nonumber\\
&&\left.-z^6\big)\big)\left(\frac{l^2}{r_+^2}\right) -\frac{4 \sqrt{3} \lambda  l^6 \Omega_H^2}{z^{14}} \Big(2 z^8 \left(3 l^2+2 z^2\right)+l^2 \Omega_H^2 \big(3 l^{12} \Omega_H^2+l^6 \left(8 z^4-2 \Omega_H^2 z^6\right)\right.\nonumber\\
&&\left.+2 z^6 \left(l^4+2 l^2 z^2-2 z^4\right)\big)\Big)+\frac{4 \sqrt{3} \lambda  l^6 \Omega_H^2}{z^{18}} \Big(10 z^{12} \left(l^2+z^2\right)+l^2 \Omega_H^2 \big(4 l^{18} \Omega_H^4+20 l^6\right.\nonumber\\
&&\left.\times  z^8+10 z^{12}\left(l^2-z^2\right)-3 l^{12} \Omega_H^2 z^4 \left(l^2 \Omega_H^2-5\right)-3 l^{10} \Omega_H^2 z^6 \left(l^2 \Omega_H^2-1\right)+2 l^4 z^{10} \right.\nonumber\\
&&\left.\times\big(l^2 \Omega_H^2 \left(l^2 \Omega_H^2-6\right)+5\big)\big)\Big)\left(\frac{r_+^2}{l^2}\right)
+\cO\left(\frac{r_+^{4}}{l^{4}}\right)\bigg)\right.\nonumber
\\
g^{in}_4(z) &=&\left(\frac{Q^4}{l^8}\right)\bigg(\frac{l^6 \Omega_H^2}{9 z^6} \left(2 \left(102 \lambda ^2+5\right) l^2+3 \left(12 \lambda ^2-1\right) z^2\right)\left(\frac{l^6}{r_+^6}\right)+\frac{\Omega_H^2}{18 z^{12} \left(l^2-z^2\right)}\nonumber\\
&&\left.\times\bigg(l^2 \left(l^2-z^2\right) \Big(18 l^{14} \Omega_H^2-24 \left(42 \lambda ^2+1\right) l^{12} \Omega_H^2 z^2+45 \left(1-12 \lambda ^2\right) l^{10} \Omega_H^2 z^4\right.\nonumber\\
&&\left.-2 \left(732 \lambda ^2+11\right) l^8 \Omega_H^2 z^6+2 l^6 z^6 \left(3 \left(96 \lambda ^2+1\right) \Omega_H^2 z^2-8 \left(102 \lambda ^2+5\right)\right)+12\right.\nonumber
\\
&&\left.\times \left(1-66 \lambda ^2\right) l^4 z^8+12 \left(1-12 \lambda ^2\right) l^2 z^{10}+144 \left(1-12 \lambda ^2\right) z^{12}\Big)+24 \left(12 \lambda ^2-1\right)\right.\nonumber\\
&&\left.\times z^{12} \Big(\left(l^4-7 l^2 z^2+6 z^4\right) \log \left(1-\frac{l^2}{z^2}\right)-2 l^2 \big(2 l^2-3 z^2\big)\Big) \log \left(\frac{z^2}{l^2}\right)-24 \big(12 \lambda ^2\right.\nonumber\\
&&\left.-1\big) z^{12} \left(l^4-7 l^2 z^2+6 z^4\right) \text{Li}_2\left(\frac{l^2}{z^2}\right)\bigg)\left(\frac{l^4}{r_+^4}\right)+\cO\left(\frac{l^{2}}{r_+^{2}}\right)\bigg)\right.\nonumber
\\
h^{in}_0(z) &=&1+\frac{l^6 \Omega_H^2}{z^4}\left(\frac{r_+^2}{l^2}\right)+\frac{l^6 \Omega_H^2 \left(l^2 \Omega_H^2-1\right)}{z^4}\left(\frac{r_+^4}{l^4}\right)+\frac{l^6 \Omega_H^2 \left(l^2 \Omega_H^2-1\right)^2}{z^4}\left(\frac{r_+^6}{l^6}\right)\nonumber\\
&&\left.+\frac{l^6 \Omega_H^2 \left(l^2 \Omega_H^2-1\right)^3}{z^4}\left(\frac{r_+^8}{l^8}\right)+\cO\left(\frac{r_+^{10}}{l^{10}}\right)\right.\nonumber
\\
h^{in}_2(z) &=&\left(\frac{Q^2}{l^4}\right)\bigg(-\frac{l^8 \Omega_H^2}{z^6}\left(\frac{l^2}{r_+^2}\right)+\frac{l^6 \Omega_H^2}{z^6} \left(l^2 \left(2-\Omega_H^2 z^2\right)+z^2\right)+\frac{l^6 \Omega_H^2}{z^6} \Big(-z^2 \big(l^2 \Omega_H^2 \nonumber\\
&&\left.\times\left(l^2 \Omega_H^2-4\right)+3\big)-3 l^2\Big)\left(\frac{r_+^2}{l^2}\right)
+\frac{l^6 \Omega_H^2}{z^6} \Big(z^2 \left(6-l^2 \Omega_H^2 \left(l^2 \Omega_H^2 \left(l^2 \Omega_H^2-5\right)+10\right)\right)\right.\nonumber\\
&&\left.+4 l^2\Big)\left(\frac{r_+^4}{l^4}\right)+\cO\left(\frac{r_+^{6}}{l^{6}}\right)\bigg)\right.\nonumber
\\
h^{in}_3(z)&=&\left(\frac{Q^3}{l^6}\right)\bigg(\frac{4 \sqrt{3} \lambda  l^8 \Omega_H^2}{z^6} \left(\frac{l^4}{r_+^4}\right)-\frac{4 \sqrt{3} \lambda  l^6 \Omega_H^2}{z^6} \left(l^2 \left(\Omega_H^2 \left(l^2-z^2\right)+3\right)+z^2\right)\left(\frac{l^2}{r_+^2}\right)\nonumber\\
&&\left.+\frac{8 \sqrt{3} \lambda  l^6 \Omega_H^2}{z^6} \left(l^2 \left(2 \Omega_H^2 \left(l^2-z^2\right)+3\right)+2 z^2\right)-\frac{40 \sqrt{3} \lambda  l^6 \Omega_H^2}{z^6}\big(l^2 \big(\Omega_H^2 \left(l^2-z^2\right)\right.\nonumber\\
&&\left.+1\big)+z^2\big)\left(\frac{r_+^2}{l^2}\right)+\cO\left(\frac{r_+^{4}}{l^{4}}\right)\bigg)\right.\nonumber
\eeqa
\beqa
h^{in}_4(z) &=&\left(\frac{Q^4}{l^8}\right)\bigg(\frac{l^6 \Omega_H^2}{3 z^6} \left(-12 \lambda ^2 \left(7 l^2+z^2\right)-2 l^2+z^2\right)\left(\frac{l^6}{r_+^6}\right)+\frac{\Omega_H^2}{18 z^{10}}\bigg(l^2 \Big(8 \left(12 \lambda ^2-1\right)\nonumber\\
&&\left.\times l^{12} \Omega_H^2+21 \left(12 \lambda ^2-1\right) l^{10} \Omega_H^2 z^2+18 \left(84 \lambda ^2+1\right) l^8 \Omega_H^2 z^4-6 l^6 z^4 \big(48 \lambda ^2 \big(2 \Omega_H^2 z^2\right.\nonumber\\
&&\left.-7\big)+\Omega_H^2 z^2-8\big)+12 \left(66 \lambda ^2-1\right) l^4 z^6+36 \left(12 \lambda ^2-1\right) l^2 z^8+144 \left(1-12 \lambda ^2\right)\right.\nonumber\\
&&\left.\times z^{10}\Big)+72 \left(12 \lambda ^2-1\right)  z^{10}\left(\left(l^2-2 z^2\right) \log \left(1-\frac{l^2}{z^2}\right)-2 l^2\right) \log \left(\frac{z^2}{l^2}\right)-72\right.\nonumber\\
&&\left.\times \left(12 \lambda ^2-1\right) z^{10} \left(l^2-2 z^2\right) \text{Li}_2\left(\frac{l^2}{z^2}\right)\bigg)\left(\frac{l^4}{r_+^4}\right)+\cO\left(\frac{l^{2}}{r_+^{2}}\right)\bigg)\right.\nonumber
\\
\Omega^{in}_0(z) &=&\frac{l^4 \Omega_H}{z^4}+\frac{l^6 \Omega_H^3}{z^8} \left(z^4-l^4\right)\left(\frac{r_+^2}{l^2}\right) +\frac{l^6 \Omega_H^3 }{z^{12}}\left(l^4-z^4\right) \left(l^2 \Omega_H^2 \left(l^4-z^4\right)+z^4\right)\left(\frac{r_+^4}{l^4}\right)\nonumber\\
&&\left.
 -\frac{l^6 \Omega_H^3 }{z^{16}}\left(l^4-z^4\right) \left(l^2 \Omega_H^2 \left(l^4-z^4\right)+z^4\right)^2\left(\frac{r_+^6}{l^6}\right) +\frac{l^6 \Omega_H^3}{z^{20}} \left(l^4-z^4\right) \big(l^2 \Omega_H^2 \left(l^4-z^4\right)\right.\nonumber\\
&&\left.+z^4\big)^3\left(\frac{r_+^8}{l^8}\right) +\cO\left(\frac{r_+^{10}}{l^{10}}\right)\right.\nonumber\\
\Omega^{in}_2(z) &=&\left(\frac{Q^2}{l^4}\right)\bigg(-\frac{l^4 \Omega_H }{z^6} \left(l^2-z^2\right)\left(\frac{l^4}{r_+^4}\right)+\frac{l^4 \Omega_H}{z^{10}} \left(l^2-z^2\right) \left(l^2 \Omega_H^2 \left(2 l^4+l^2 z^2+z^4\right)+z^4\right)\nonumber\\
&&\left.\times\left(\frac{l^2}{r_+^2}\right) -\frac{l^4 \Omega_H }{z^{14}}\left(l^2-z^2\right) \Big(z^8 \left(1-l^2 \Omega_H^2 \left(l^2 \Omega_H^2-3\right)\right)+l^4 \Omega_H^2 z^6 \left(3-l^2 \Omega_H^2\right)+l^6 \right.\nonumber\\
&&\left.\times\Omega_H^2 \left(l^4 \Omega_H^2 \left(3 l^2+2 z^2\right)+4 z^4\right)\Big)+\frac{l^4 \Omega_H}{z^{18}} \big(-z^{14} \big(l^2 \Omega_H^2 \left(l^2 \Omega_H^2 \left(l^2 \Omega_H^2-4\right)+6\right)\right.\nonumber\\
&&\left.+1\big)+l^2 z^{12}+l^{18} \Omega_H^6 \left(4 l^2-z^2\right)+3 l^{14} \Omega_H^4 z^4 \left(3-2 l^2 \Omega_H^2\right)-l^{12} \Omega_H^4 z^6 \left(l^2 \Omega_H^2+1\right)\right.\nonumber\\
&&\left.+2 l^8 \Omega_H^2 z^8 \left(l^2 \Omega_H^2 \left(l^2 \Omega_H^2-3\right)+3\right)+3 l^8 \Omega_H^4 z^{10} \left(l^2 \Omega_H^2-2\right)\big)\left(\frac{r_+^2}{l^2}\right) +\frac{l^4 \Omega_H }{z^{22}} \Big(z^{18} \right.\nonumber\\
&&\left.\times\big(1-l^2 \Omega_H^2 \left(l^2 \Omega_H^2 \left(l^2 \Omega_H^2 \left(l^2 \Omega_H^2-5\right)+10\right)-10\right)\big)-l^2 z^{16}+l^{24} \Omega_H^8 \left(z^2-5 l^2\right)\right.\nonumber\\
&&\left.+4 l^{20} \Omega_H^6 z^4 \left(3 l^2 \Omega_H^2-4\right)+l^{18} \Omega_H^6 z^6 \left(l^2 \Omega_H^2+1\right)-3 l^{14} \Omega_H^4 z^8 \big(l^2 \Omega_H^2 \left(3 l^2 \Omega_H^2-8\right)\right.\nonumber\\
&&\left.+6\big)+l^{12} \Omega_H^4 z^{10}  \left(-l^2 \Omega_H^2 \left(6 l^2 \Omega_H^2-11\right)-2\right)+2 l^8 \Omega_H^2 z^{12} \big(l^2 \Omega_H^2 \big(l^2 \Omega_H^2 \left(l^2 \Omega_H^2-4\right)\right.\nonumber\\
&&\left.+6\big)-4\big)+l^6 \Omega_H^2 z^{14} \left(l^2 \Omega_H^2 \left(l^2 \Omega_H^2 \left(5 l^2 \Omega_H^2-17\right)+18\right)-2\right)\Big)\left(\frac{r_+^4}{l^4}\right)
+\cO\left(\frac{r_+^{6}}{l^{6}}\right)\bigg)\right.\nonumber
\\
\Omega^{in}_3(z) &=&\left(\frac{Q^3}{l^6}\right)\bigg(\frac{2 \sqrt{3} \lambda  l^4 \Omega_H}{z^6} \left(l^2-z^2\right)\left(\frac{l^6}{r_+^6}\right)-\frac{2 \sqrt{3} \lambda  l^4 \Omega_H}{z^{10}} \left(l^2-z^2\right) \Big(z^4 \left(3 l^2 \Omega_H^2+2\right)\nonumber\\
&&\left. +l^4 \Omega_H^2\left(3 l^2+2 z^2\right)\Big)\left(\frac{l^4}{r_+^4}\right) +\frac{2 \sqrt{3} \lambda  l^4 \Omega_H}{z^{14}} \left(l^2-z^2\right) \Big(z^8 \left(11 l^2 \Omega_H^2+3\right)+l^6 \Omega_H^2 z^4 \right.\nonumber\\
&&\left.\times\big(4 l^2  \Omega_H^2+9\big)+l^4 \Omega_H^2 \left(l^6 \Omega_H^2 \left(5 l^2+4 z^2\right)+8 z^6\right)\Big)\left(\frac{l^2}{r_+^2}\right) -\frac{2 \sqrt{3} \lambda  l^4 \Omega_H}{z^{18}} \left(l^2-z^2\right)\right.\nonumber\\
&&\left. \times \Big(20 l^4 \Omega_H^2 z^{10}+2 z^{12} \left(13 l^2 \Omega_H^2+2\right)+l^{16} \Omega_H^6 \left(7 l^2+6 z^2\right)+l^{12} \Omega_H^4 z^4 \left(l^2 \Omega_H^2+20\right)\right.\nonumber
\\
&&\left.-4 l^{10} \Omega_H^4 z^6 \left(l^2 \Omega_H^2-5\right)+2 l^6 \Omega_H^2 z^8 \left(9-2 l^2 \Omega_H^2 \left(l^2 \Omega_H^2-5\right)\right)\Big)+\frac{2 \sqrt{3} \lambda  l^4 \Omega_H}{z^{22}} \big(l^2\right.\nonumber
\eeqa
\beqa
&&\left.-z^2\big)\Big(40 l^4 \Omega_H^2 z^{14}+5 z^{16} \left(10 l^2 \Omega_H^2+1\right)+l^{22} \Omega_H^8 \left(9 l^2+8 z^2\right)+l^{18} \Omega_H^6 z^4 \big(35-6 l^2\right.\nonumber\\
&&\left.\times \Omega_H^2\big)-12 l^{16} \Omega_H^6 z^6 \left(l^2 \Omega_H^2-3\right)+l^{12} \Omega_H^4 z^8 \left(50-l^2 \Omega_H^2 \left(7 l^2 \Omega_H^2-11\right)\right)+4 l^{10} \Omega_H^4 \right.\nonumber\\
&&\left. \times z^{10} \big(l^2 \Omega_H^2 \left(l^2 \Omega_H^2-6\right)+15\big)+2 l^6 \Omega_H^2 z^{12}\left(2 l^2 \Omega_H^2 \left(l^2 \Omega_H^2 \left(l^2 \Omega_H^2-6\right)+15\right)+15\right)\Big)\right.\nonumber\\
&&\left.\times\left(\frac{r_+^2}{l^2}\right)
+\cO\left(\frac{r_+^{4}}{l^{4}}\right)\bigg)\right.\nonumber
\eeqa
\beqa
\Omega^{in}_4(z) &=&\left(\frac{Q^4}{l^8}\right)\bigg(\frac{12 \lambda ^2 l^4 \Omega_H}{z^6} \left(z^2-l^2\right)\left(\frac{l^8}{r_+^8}\right)-\frac{l^4 \Omega_H}{9 z^{12}} \left(l^2-z^2\right)\Big(9 l^8 \Omega_H^2-2 \left(204 \lambda ^2+1\right) l^6 \nonumber\\
&&\left.\times \Omega_H^2 z^2+4 \left(1-93 \lambda ^2\right) l^4 \Omega_H^2 z^4+\left(1-660 \lambda ^2\right) l^2 \Omega_H^2 z^6-324 \lambda ^2 z^6\Big)\left(\frac{l^6}{r_+^6}\right)\right.\nonumber\\
&&\left.+\frac{l^2 \Omega_H}{18 z^{16}}\bigg(l^2 \Big(54 l^{16} \Omega_H^4-4 \left(354 \lambda ^2+11\right) l^{14} \Omega_H^4 z^2+2 \left(17-96 \lambda ^2\right) l^{12} \Omega_H^4 z^4\right.\nonumber\\
&&\left.+l^{10} \Omega_H^2 z^4 \big(54-\left(1356 \lambda ^2+49\right) \Omega_H^2 z^2\big)+l^8 \Omega_H^2 z^6 \Big(\left(1716 \lambda ^2+19\right) \Omega_H^2 z^2\right.\nonumber
\\
&&\left.-4 \left(816 \lambda ^2+13\right)\Big)-24 l^6 \Omega_H^2 z^8 \left(6 \lambda ^2 \left(3 \Omega_H^2 z^2+2\right)-1\right)
+2 l^4 \Omega_H^2 z^{10} \big(-1428 \lambda ^2\right.\nonumber\\
&&\left.+7 \left(120 \lambda ^2-1\right) \Omega_H^2 z^2+11\big)+8 l^2 z^{10} \Big(\left(3 \left(315+4 \pi ^2\right) \lambda ^2-\pi ^2-18\right) \Omega_H^2 z^2-162 \lambda ^2\Big)\right.\nonumber\\
&&\left.+48 z^{12} \big(27 \lambda ^2+2 \big(1-12 \lambda ^2\big) \Omega_H^2 z^2\big)\Big)-48 \left(12 \lambda ^2-1\right) \Omega_H^2 z^{12} \left(z^2-2 l^2\right)\Big(2 l^2-\left(l^2-2 z^2\right)\right.\nonumber\\
&&\left. \times \log \left(1-\frac{l^2}{z^2}\right)\Big) \log \left(\frac{z^2}{l^2}\right)+48 \left(12 \lambda ^2-1\right) \Omega_H^2 z^{12} \left(2 l^4-5 l^2 z^2+2 z^4\right)  \text{Li}_2\left(\frac{l^2}{z^2}\right)\bigg)\right.\nonumber\\
&&\left.\times\left(\frac{l^4}{r_+^4}\right)+\cO\left(\frac{l^{2}}{r_+^{2}}\right)\bigg) \right.\nonumber
\eeqa
and also
\beqa
{a^{in}_0}_1(z) &=&\left(\frac{Q}{l^2}\right)\left(\frac{\sqrt{3} l^2}{z^2}\left(\frac{l}{r_+}\right)+\cO\left(\frac{r_+^{11}}{l^{11}}\right)\right)\nonumber
\\
{a^{in}_0}_2(z)&=&\left(\frac{Q^2}{l^4}\right)\left(\cO\left(\frac{r_+^{7}}{l^{7}}\right)\right)
\label{a0inq2}
\\
{a^{in}_0}_3(z) &=&\left(\frac{Q^3}{l^6}\right)\bigg(-\frac{\left(12 \lambda ^2-1\right) l^8 \Omega_H^2}{\sqrt{3} z^6}\left(\frac{l^3}{r_+^3}\right)-\frac{\left(12 \lambda ^2-1\right) l^8 \Omega_H^2}{3 \sqrt{3} z^8} \left(l^2 \Omega_H^2 \left(2 l^2-z^2\right)-9 z^2\right)\nonumber\\
&&\left.\times\left(\frac{l}{r_+}\right)+\frac{\left(12 \lambda ^2-1\right) l^4 \Omega_H^2}{6 \sqrt{3} z^{10}}\bigg(-3 l^{12} \Omega_H^4-8 l^2 \Omega_H^2 z^6 \left(l^2-4 z^2\right)-2 l^8 \Omega_H^2 z^2 \big(l^2 \Omega_H^2\right.\nonumber\\
&&\left.-10\big)+2 l^4 z^4 \left(l^2 \Omega_H^2-6\right) \left(l^2 \Omega_H^2+3\right)+16 \Omega_H^2 z^8 \left(2 l^2-\left(l^2-2 z^2\right) \log \left(1-\frac{l^2}{z^2}\right)\right)\right.\nonumber\\
&&\left.\times \log \left(\frac{z^2}{l^2}\right)+16 \Omega_H^2 z^8 \left(l^2-2 z^2\right) \text{Li}_2\left(\frac{l^2}{z^2}\right)\bigg)\left(\frac{r_+}{l}\right) +\cO\left(\frac{r_+^{3}}{l^{3}}\right)\bigg)\right.\nonumber
\eeqa
\beqa
{a^{in}_0}_4(z) &=&\left(\frac{Q^4}{l^8}\right)\bigg(\frac{4 \lambda  \left(12 \lambda ^2-1\right) l^8 \Omega_H^2}{z^6}\left(\frac{l}{r_+}\right)^5+\frac{8 \lambda  \left(12 \lambda ^2-1\right) l^8 \Omega_H^2 }{3 z^8}\Big(l^4 \Omega_H^2-2 z^2 \big(l^2 \Omega_H^2\nonumber\\
&&\left.+3\big)\Big)\left(\frac{l^3}{r_+^3}\right)+\cO \left(\frac{l}{r_+}\right)\bigg)\right.\nonumber
\\
{a^{in}_1}_1(z) &=&\left(\frac{Q}{l^2}\right)\left(\frac{\sqrt{3} l^4 \Omega_H}{z^2}\left(-1 + \left(\frac{r_+^2}{l^2}\right) - \left(\frac{r_+^4}{l^4}\right) + \left(\frac{r_+^6}{l^6}\right) - \left(\frac{r_+^8}{l^8}\right)\right)+\cO\left(\frac{r_+^{10}}{l^{10}}\right)\right)\nonumber
\\
{a^{in}_1}_2(z) &=&\left(\frac{Q^2}{l^4}\right)\bigg(\frac{6 \lambda  l^4 \Omega_H}{z^2}\left(\frac{l^2}{r_+^2}\right)-\frac{6 \lambda  l^4 \Omega_H \left(l^2 \Omega_H^2+2\right)}{z^2}+\frac{18 \lambda  l^4 \Omega_H \left(l^2 \Omega_H^2+1\right)}{z^2}\left(\frac{r_+^2}{l^2}\right)\nonumber\\
&&\left.-\frac{12 \lambda  l^4 \Omega_H \left(3 l^2 \Omega_H^2+2\right)}{z^2}\left(\frac{r_+^4}{l^4}\right)+\cO\left(\frac{r_+^{6}}{l^{6}}\right)\bigg)\right.\nonumber
\\
{a^{in}_1}_3(z) &=&\left(\frac{Q^3}{l^6}\right)\bigg(-\frac{12 \sqrt{3} \lambda ^2 l^4 \Omega_H}{z^2}\left(\frac{l^4}{r_+^4}\right)+\frac{l^4 \Omega_H}{3 \sqrt{3} z^6}\Big(3 \left(12 \lambda ^2-1\right) l^6 \Omega_H^2+4 \left(12 \lambda ^2-1\right) l^4 \nonumber
\\
&&\left.\times\Omega_H^2 z^2+3 \left(96 \lambda ^2+1\right) l^2 \Omega_H^2 z^4+324 \lambda ^2 z^4\Big)\left(\frac{l^2}{r_+^2}\right)+\frac{l^2 \Omega_H}{6 \sqrt{3} z^8}\bigg(l^2 \Big(4 \left(12 \lambda ^2-1\right) l^{10}\right.\nonumber
\\
&&\left.\times \Omega_H^4+3 \big(12 \lambda ^2
-1\big) l^8 \Omega_H^4 z^2-4 \left(12 \lambda ^2-1\right) l^6 \Omega_H^2 z^2 \left(\Omega_H^2 z^2+6\right)-8 l^4 \Omega_H^2 z^4 \big(6 \lambda ^2\right.\nonumber\\
&&\left.\times \left(9 \Omega_H^2 z^2+10\right)-5\big)-18 \left(132 \lambda ^2+1\right) l^2 \Omega_H^2 z^6+16 \Omega_H^2 z^8-48 \lambda ^2 z^6 \big(4 \Omega_H^2 z^2\right.\nonumber\\
&&\left.+27\big)\Big)-16 \left(12 \lambda ^2-1\right) \Omega_H^2 z^8 \left(z^2 \log \left(1-\frac{l^2}{z^2}\right)+l^2\right) \log \left(\frac{z^2}{l^2}\right)+16 \big(12 \lambda ^2\right.\nonumber\\
&&\left.-1\big) \Omega_H^2 z^{10} \text{Li}_2\left(\frac{l^2}{z^2}\right)\bigg)+\cO\left(\frac{r_+^{2}}{l^{2}}\right)\bigg)\right.\nonumber
\\
{a^{in}_1}_4(z) &=&\left(\frac{Q^4}{l^8}\right)\bigg(\frac{72 \lambda ^3 l^4 \Omega_H}{z^2}\left(\frac{l^6}{r_+^6}\right)-\frac{2 \lambda  l^4 \Omega_H }{3 z^6} \Big(8 \left(12 \lambda ^2-1\right) l^6 \Omega_H^2+12 \left(12 \lambda ^2-1\right)\nonumber\\
&&\left.\times l^4 \Omega_H^2 z^2+3 \left(204 \lambda ^2+1\right) l^2 \Omega_H^2 z^4+432 \lambda ^2 z^4\Big)\left(\frac{l^4}{r_+^4}\right)+\cO\left(\frac{l^{2}}{r_+^{2}}\right)\bigg)\right.\nonumber
\eeqa
For the gauge field.

 It is straightforward to check that for $\lambda=\lambda_{SG}$  the expansions in section \ref{NFP} for the  supergravity solution are
recovered.

%\\
%And for the inner solution, we get
%\\
%From the Maxwell-Chern-Simons equation, we find the radial form of equations and use the described matching method
%\\
%the above relations are equal to the exact supergravity solution at linear order in Q.
\section{Expansion of the Supergravity Solution} \label{pertpope}

 In this appendix we present for comparison and completeness the far-field and near-field expansions of
the exact solution \eqref{PopeKAdS}.

\subsection{Far-field Expansion}\label{FFP}

The far-field expansion of  \eqref{PopeKAdS} is
\beqa
f_{SG}^{out}(r)&=&1+\frac{r^2}{l^2}-\frac{l^2}{r^2}\left(\frac{r_+^2}{l^2}\right)-\frac{l^2 \left(l^2 \Omega_H^2+1\right)}{r^2}\left(\frac{r_+^4}{l^4}\right) +\frac{l^4 \Omega_H^2}{r^4} \left(l^2 \left(1-\Omega_H^2 r^2\right)+r^2\right)\left(\frac{r_+^6}{l^6}\right)\nonumber\\
&&\left.-\frac{l^4 \Omega_H^2}{r^4} \left(l^2 \Omega_H^2-1\right) \left(l^2 \left(\Omega_H^2 r^2-1\right)-r^2\right)\left(\frac{r_+^8}{l^8}\right) +\cO\left(\frac{r_+^{10}}{l^{10}}\right) +\left(\frac{Q^2}{l^4}\right)\bigg(-\frac{l^2}{r^2}\right.\nonumber\\
&&\left.\times\left(\frac{l^2}{r_+^2}\right)+\frac{l^4}{r^4}+\frac{l^6 \Omega_H^4}{r^2}\left(\frac{r_+^2}{l^2}\right) +\frac{l^4 \Omega_H^2}{r^4} \left(l^2 \Omega_H^2 \left(l^2 \left(\Omega_H^2 r^2-1\right)-4 r^2\right)+r^2\right)\left(\frac{r_+^4}{l^4}\right)\right.\nonumber\\
&&\left.+\cO\left(\frac{r_+^{6}}{l^{6}}\right)\bigg)+\left(\frac{Q^3}{l^6}\right)\bigg(-\frac{2 l^6 \Omega_H^4}{r^2}+\frac{2 l^4 \Omega_H^2}{r^4} \left(l^2 \Omega_H^2 \left(l^2+5 r^2\right)-r^2\right)\left(\frac{r_+^2}{l^2}\right)\right.\nonumber\\
&&\left.+\cO\left(\frac{r_+^{4}}{l^{4}}\right)\bigg)
+\left(\frac{Q^4}{l^8}\right)\left(\cO\left(\frac{l^{2}}{r_+^{2}}\right)\right)\right.
\nonumber
\\
g_{SG}^{out}(r)&=&1-\frac{l^6 \Omega_H^2}{r^4}\left(\frac{r_+^6}{l^6}\right)-\frac{l^6 \Omega_H^2 \left(l^2 \Omega_H^2-1\right)}{r^4}\left(\frac{r_+^8}{l^8}\right)+\cO\left(\frac{r_+^{10}}{l^{10}}\right)
+\left(\frac{Q^2}{l^4}\right)\bigg(\frac{l^6 \Omega_H^2}{r^6} \big(r^2 \nonumber\\
&&\left.\times\left(l^2 \Omega_H^2-1\right)+l^2\big)\left(\frac{r_+^4}{l^4}\right)+\cO\left(\frac{r_+^{6}}{l^{6}}\right)\bigg)+\left(\frac{Q^3}{l^6}\right)\bigg(-\frac{2 l^6 \Omega_H^2}{r^6} \big(r^2 \left(l^2 \Omega_H^2-1\right)\right.\nonumber\\
&&\left.+l^2\big)\left(\frac{r_+^2}{l^2}\right)+\cO\left(\frac{r_+^{4}}{l^{4}}\right)\bigg)
+\left(\frac{Q^4}{l^8}\right)\left(\cO\left(\frac{l^{2}}{r_+^{2}}\right)\right) \right.\nonumber
\\
h_{SG}^{out}(r)&=&1+\frac{l^6 \Omega_H^2}{r^4}\left(\frac{r_+^6}{l^6}\right)+\frac{l^6 \Omega_H^2 \left(l^2 \Omega_H^2-1\right)}{r^4}\left(\frac{r_+^8}{l^8}\right)+\cO\left(\frac{r_+^{10}}{l^{10}}\right) +\left(\frac{Q^2}{l^4}\right)\bigg(-\frac{l^6 \Omega_H^2}{r^6}\big(r^2\nonumber\\
&&\left.\times \big(l^2 \Omega_H^2-1\big)+l^2\big)\left(\frac{r_+^4}{l^4}\right)
+\cO\left(\frac{r_+^{6}}{l^{6}}\right)\bigg)+\left(\frac{Q^3}{l^6}\right)\bigg(\frac{2 l^6 \Omega_H^2}{r^6} \left(r^2 \left(l^2 \Omega_H^2-1\right)+l^2\right)\right.\nonumber\\
&&\left.\times\left(\frac{r_+^2}{l^2}\right)
+\cO\left(\frac{r_+^{4}}{l^{4}}\right)\bigg)+\left(\frac{Q^4}{l^8}\right)\left(\cO\left(\frac{l^{2}}{r_+^{2}}\right)\right)\right.\nonumber
\\
\Omega_{SG}^{out}(r)&=&\frac{l^4 \Omega_H}{r^4}\left(\frac{r_+^4}{l^4}\right)+\frac{l^6 \Omega_H^3}{r^4}\left(\frac{r_+^6}{l^6}\right)+\frac{l^6 \Omega_H^3 \left(l^2 \Omega_H^2-1\right)}{r^4}\left(\frac{r_+^8}{l^8}\right)+\cO\left(\frac{r_+^{10}}{l^{10}}\right)
+\left(\frac{Q^2}{l^4}\right)\nonumber\\
&&\left.\times\bigg(\frac{l^4 \Omega_H}{r^4}-\frac{l^4 \Omega_H }{r^6}\left(l^2 \left(\Omega_H^2 r^2+1\right)+r^2\right)\left(\frac{r_+^2}{l^2}\right)+\frac{l^4 \Omega_H}{r^6} \big(r^2 \left(1-l^2 \Omega_H^2 \left(l^2 \Omega_H^2-3\right)\right)\right.\nonumber\\
&&\left.+l^2\big)\left(\frac{r_+^4}{l^4}\right)+\cO\left(\frac{r_+^{6}}{l^{6}}\right)
\bigg)+\left(\frac{Q^3}{l^6}\right)\bigg(-\frac{l^4 \Omega_H}{r^4}\left(\frac{l^2}{r_+^2}\right)+\frac{l^4 \Omega_H }{r^6} \big(r^2 \left(3 l^2 \Omega_H^2+2\right)\right.\nonumber\\
&&\left. +l^2\big)-\frac{l^4 \Omega_H}{r^6}\left(l^2 \left(\Omega_H^2 \left(l^2+11 r^2\right)+2\right)+3 r^2\right)\left(\frac{r_+^2}{l^2}\right)+\cO\left(\frac{r_+^{4}}{l^{4}}\right)\bigg)+\left(\frac{Q^4}{l^8}\right)\right.\nonumber\\
&&\left.\times\left(\frac{l^4 \Omega_H}{r^4}\left(\frac{l^4}{r_+^4}\right)+\cO\left(\frac{l^{2}}{r_+^{2}}\right)\right)\right.\nonumber
\eeqa
for the metric functions.  For the gauge field we obtain
\beqa
{a_0^{out}}_{SG}(r)&=&\left(\frac{Q}{l^2}\right)\frac{\sqrt{3} l^2 }{r^2}\nonumber
\\
{a_1^{out}}_{SG}&=&\left(\frac{Q}{l^2}\right)\left(\frac{\sqrt{3} l^4 \Omega_H}{r^2}\left(-\left(\frac{r_+^2}{l^2}\right) + \left(\frac{r_+^4}{l^4}\right) - \left(\frac{r_+^6}{l^6}\right) + \left(\frac{r_+^8}{l^8}\right)\right) +\cO\left(\frac{r_+^{10}}{l^{10}}\right)\right)
\nonumber\\
&&+\left(\frac{Q^2}{l^4}\right)\bigg(\frac{\sqrt{3} l^4 \Omega_H}{r^2}
\left.-\frac{\sqrt{3} l^4 \Omega_H \left(l^2 \Omega_H^2+2\right)}{r^2}\left(\frac{r_+^2}{l^2}\right) +\frac{3 \sqrt{3} l^4 \Omega_H \left(l^2 \Omega_H^2+1\right)}{r^2}\right.\nonumber\\
&&\left.\times\left(\frac{r_+^4}{l^4}\right)+\cO\left(\frac{r_+^{6}}{l^{6}}\right)
\bigg)  +\left(\frac{Q^3}{l^6}\right) \bigg(-\frac{\sqrt{3} l^4 \Omega_H}{r^2}\left(\frac{l^2}{r_+^2}\right)+\frac{3 \sqrt{3} l^4 \Omega_H \left(l^2 \Omega_H^2+1\right)}{r^2}\right.\nonumber\\
&&\left.-\frac{2 \sqrt{3} l^4 \Omega_H }{r^2}\left(l^2 \Omega_H^2 \left(l^2 \Omega_H^2+6\right)+3\right)\left(\frac{r_+^2}{l^2}\right)+\cO\left(\frac{r_+^{4}}{l^{4}}\right)
\bigg)+\left(\frac{Q^4}{l^8}\right)\bigg(\frac{\sqrt{3} l^4 \Omega_H}{r^2}\right.\nonumber\\
&&\left.\times\left(\frac{l^4}{r_+^4}\right)+\cO\left(\frac{l^{2}}{r_+^{2}}\right)
\bigg)\right.\nonumber
\eeqa

\subsection{Near-field Expansion}
\label{NFP}

The near-field expansion of \eqref{PopeKAdS} is considerably lengthier, and so we present the results for
the metric functions in successive powers of $Q/l^2$.  The index represents the order of $Q/l^2$.  We find
\beqa
{f^{in}_{SG}}_0(z)&=&1-\frac{l^2}{z^2}+\frac{1}{l^2 z^4}\left(l^8 \Omega_H^2-l^4 z^2 \left(l^2 \Omega_H^2+1\right)+z^6\right)\left(\frac{r_+^2}{l^2}\right) +\frac{l^4 \Omega_H^2}{z^4} \left(l^2 \Omega_H^2-1\right) \nonumber\\
&&\left.\times\left(l^2 -z^2\right)\left(\frac{r_+^4}{l^4}\right)+\frac{l^4 \Omega_H^2}{z^4} \left(l^2 \Omega_H^2-1\right)^2 \left(l^2-z^2\right)\left(\frac{r_+^6}{l^6}\right) +\frac{l^4 \Omega_H^2}{z^4} \left(l^2 \Omega_H^2-1\right)^3 \right.\nonumber\\
&&\left.\times\left(l^2-z^2\right)\left(\frac{r_+^8}{l^8}\right)
+\cO\left(\frac{r_+^{10}}{l^{10}}\right)\right.\nonumber
\\
{f^{in}_{SG}}_2(z)&=&\left(\frac{Q^2}{l^4}\right)\bigg(\frac{l^2 \left(l^2-z^2\right)}{z^4}\left(\frac{l^4}{r_+^4}\right)-\frac{l^6 \Omega_H^4 \left(l^2-z^2\right)}{z^4}-\frac{l^4 \Omega_H^2}{z^4} \left(l^2 \Omega_H^2 \left(l^2 \Omega_H^2-4\right)+1\right) \nonumber\\
&&\left. \left(l^2-z^2\right)\left(\frac{r_+^2}{l^2}\right) -\frac{l^4 \Omega_H^2 }{z^4}\left(l^2 \Omega_H^2 \left(l^2 \Omega_H^2 \left(l^2 \Omega_H^2-5\right)+10\right)-3\right) \left(l^2-z^2\right)\left(\frac{r_+^4}{l^4}\right)\right.\nonumber\\
&&\left.+\cO\left(\frac{r_+^{6}}{l^{6}}\right)\bigg)\right.\nonumber
\\
{f^{in}_{SG}}_3(z)&=&\left(\frac{Q^3}{l^6}\right)\bigg(\frac{2 l^6 \Omega_H^4}{z^4} \left(l^2-z^2\right)\left(\frac{l^2}{r_+^2}\right) +\frac{2 l^4 \Omega_H^2}{z^4} \left(1-5 l^2 \Omega_H^2\right) \left(l^2-z^2\right)+\frac{2 l^4 \Omega_H^2 }{z^4}\nonumber\\
&&\left. \times\left(14 l^2 \Omega_H^2-4\right) \left(l^2-z^2\right)\left(\frac{r_+^2}{l^2}\right)+\cO\left(\frac{r_+^{4}}{l^{4}}\right)\bigg) \right.\nonumber
\eeqa
\beqa
{f^{in}_{SG}}_4(z)&=&\left(\frac{Q^4}{l^8}\right)\bigg(\frac{3 l^6 \Omega_H^4 \left(z^2-l^2\right)}{z^4}\left(\frac{l^4}{r_+^4}\right)+\cO\left(\frac{l^{2}}{r_+^{2}}\right)\bigg)
\\
{g^{in}_{SG}}_0(z)&=&1-\frac{l^6 \Omega_H^2}{z^4}\left(\frac{r_+^2}{l^2}\right) +\frac{l^6 \Omega_H^2 }{z^8} \left(l^2 \Omega_H^2 \left(l^4-z^4\right)+z^4\right)\left(\frac{r_+^4}{l^4}\right) -\frac{l^6 \Omega_H^2}{z^{12}} \big(l^2 \Omega_H^2 \left(l^4-z^4\right)\nonumber\\
&&\left.+z^4\big)^2\left(\frac{r_+^6}{l^6}\right) +\frac{l^6 \Omega_H^2}{z^{16}} \left(l^2 \Omega_H^2 \left(l^4-z^4\right)+z^4\right)^3\left(\frac{r_+^8}{l^8}\right)+\cO\left(\frac{r_+^{10}}{l^{10}}\right)\right.\nonumber
\\
{g^{in}_{SG}}_2(z)&=&\left(\frac{Q^2}{l^4}\right)\bigg(\frac{l^8 \Omega_H^2}{z^6}\left(\frac{l^2}{r_+^2}\right) -\frac{l^6 \Omega_H^2}{z^{10}} \left(2 l^8 \Omega_H^2+l^2 z^4 \left(2-\Omega_H^2 z^2\right)+z^6\right)+\frac{l^6 \Omega_H^2 }{z^{14}} \Big(3 l^{14} \Omega_H^4\nonumber\\
&&\left.+z^{10} \left(l^2 \Omega_H^2 \left(l^2 \Omega_H^2-4\right)+3\right)+3 l^2 z^8-2 l^8 \Omega_H^2 z^4 \left(l^2 \Omega_H^2-3\right)-2 l^6 \Omega_H^2 z^6 \big(l^2 \Omega_H^2\right.\nonumber\\
&&\left.-1\big)\Big)\left(\frac{r_+^2}{l^2}\right)
+\frac{l^6 \Omega_H^2}{z^{18}} \big(-4 l^{20} \Omega_H^6-4 l^2 z^{12}+6 l^{14} \Omega_H^4 z^4 \left(l^2 \Omega_H^2-2\right)+3 l^{12} \Omega_H^4 z^6 \right.\nonumber\\
&&\left.\times\left(l^2 \Omega_H^2-1\right) -2 l^8 \Omega_H^2 z^8\left(l^2 \Omega_H^2 \left(l^2 \Omega_H^2-4\right)+6\right)-4 l^6 \Omega_H^2 z^{10} \big(l^2 \Omega_H^2 \left(l^2 \Omega_H^2-3\right)\right.\nonumber\\
&&\left.+2\big)+z^{14} \big(l^2 \Omega_H^2 \left(l^4 \Omega_H^4-5 l^2 \Omega_H^2+10\right)-6\big)\Big)\left(\frac{r_+^4}{l^4}\right)+\cO\left(\frac{r_+^{6}}{l^{6}}\right)\bigg)\right.\nonumber
\\
{g^{in}_{SG}}_3(z)&=&\left(\frac{Q^3}{l^6}\right)\bigg(-\frac{2 l^8 \Omega_H^2}{z^6}\left(\frac{l^4}{r_+^4}\right) +\frac{2 l^6 \Omega_H^2}{z^{10}} \left(l^4 \Omega_H^2 \left(2 l^4+z^4\right)+l^2 z^4 \left(3-\Omega_H^2 z^2\right)+z^6\right)\nonumber\\
&&\left.\times\left(\frac{l^2}{r_+^2}\right)
-\frac{2 l^6 \Omega_H^2}{z^{14}} \Big(-4 z^{10} \left(l^2 \Omega_H^2-1\right)+2 l^2 z^8 \left(2 l^2 \Omega_H^2+3\right)+l^8 \Omega_H^2 \big(3 l^6 \Omega_H^2\right.\nonumber\\
&&\left.+8 z^4\big)-2 l^6 \Omega_H^2 z^6 \left(l^2 \Omega_H^2-1\right)\Big)+\frac{2 l^6 \Omega_H^2 }{z^{18}} \Big(4 l^{20} \Omega_H^6+20 l^8 \Omega_H^2 z^8-10 z^{14} \big(l^2 \Omega_H^2\right.\nonumber
\\
&&\left.-1\big)+10 l^2 z^{12} \left(l^2 \Omega_H^2+1\right)
-3 l^{14} \Omega_H^4 z^4 \left(l^2 \Omega_H^2-5\right)-3 l^{12} \Omega_H^4 z^6 \left(l^2 \Omega_H^2-1\right)\right.\nonumber\\
&&\left.+2 l^6 \Omega_H^2 z^{10} \left(l^2 \Omega_H^2 \left(l^2 \Omega_H^2-6\right)+5\right)\Big)\left(\frac{r_+^2}{l^2}\right)+\cO\left(\frac{r_+^{4}}{l^{4}}\right)\bigg)\right.\nonumber
\\
{g^{in}_{SG}}_4(z)&=&\left(\frac{Q^4}{l^8}\right)\bigg(\frac{3 l^8 \Omega_H^2}{z^6}\left(\frac{l^6}{r_+^6}\right) +\frac{l^6 \Omega_H^2 }{z^{12}} \Big(3 z^8 \left(l^2 \Omega_H^2-1\right)-4 l^2 z^6 \left(2 l^2 \Omega_H^2+3\right)+l^8 \Omega_H^2\nonumber\\
&&\left.\times \left(l^2-6 z^2\right)\Big)\left(\frac{l^4}{r_+^4}\right)+\cO\left(\frac{l^{2}}{r_+^{2}}\right)\bigg)\right.\nonumber
\\
{h^{in}_{SG}}_0(z)&=&1+\frac{l^6 \Omega_H^2}{z^4}\left(\frac{r_+^2}{l^2}\right)+\frac{l^6 \Omega_H^2}{z^4} \left(l^2 \Omega_H^2-1\right)\left(\frac{r_+^4}{l^4}\right) +\frac{l^6 \Omega_H^2}{z^4} \left(l^2 \Omega_H^2-1\right)^2\left(\frac{r_+^6}{l^6}\right) \nonumber\\
&&\left.+\frac{l^6 \Omega_H^2}{z^4}\left(l^2 \Omega_H^2-1\right)^3\left(\frac{r_+^8}{l^8}\right)+\cO\left(\frac{r_+^{10}}{l^{10}}\right)\right.\nonumber
\\
{h^{in}_{SG}}_2(z)&=&\left(\frac{Q^2}{l^4}\right)\bigg(-\frac{l^8 \Omega_H^2}{z^6}\left(\frac{l^2}{r_+^2}\right)+\frac{l^6 \Omega_H^2}{z^6} \left(l^2 \left(2-\Omega_H^2 z^2\right)+z^2\right)-\frac{l^6 \Omega_H^2}{z^6} \big(z^2 \big(l^2 \Omega_H^2 \nonumber\\
&&\left.\times\left(l^2 \Omega_H^2-4\right)+3\big)+3 l^2\big)\left(\frac{r_+^2}{l^2}\right) +\frac{l^6 \Omega_H^2}{z^6} \Big(z^2 \big(6-l^2 \Omega_H^2 \big(l^2 \Omega_H^2 \left(l^2 \Omega_H^2-5\right)\right.\nonumber
\\
&&\left.+10\big)\big)+4 l^2\Big)\left(\frac{r_+^4}{l^4}\right)+\cO\left(\frac{r_+^{6}}{l^{6}}\right)
\bigg)\right.\nonumber
\eeqa
\beqa
{h^{in}_{SG}}_3(z)&=&\left(\frac{Q^3}{l^6}\right)\bigg(\frac{2 l^8 \Omega_H^2}{z^6}\left(\frac{l^4}{r_+^4}\right)-\frac{2 l^6 \Omega_H^2 }{z^6} \left(l^2 \left(\Omega_H^2 \left(l^2-z^2\right)+3\right)+z^2\right)\left(\frac{l^2}{r_+^2}\right) +\frac{4 l^6 \Omega_H^2}{z^6} \nonumber\\
&&\left.\times\left(l^2 \left(2 \Omega_H^2 \left(l^2-z^2\right)+3\right)+2 z^2\right)-\frac{20 l^6 \Omega_H^2}{z^6} \left(l^2 \left(\Omega_H^2 \left(l^2-z^2\right)+1\right)+z^2\right)\left(\frac{r_+^2}{l^2}\right)
\right.\nonumber\\
&&\left.+\cO\left(\frac{r_+^{4}}{l^{4}}\right)\bigg)\right.\nonumber
\\
{h^{in}_{SG}}_4(z)&=&\left(\frac{Q^4}{l^8}\right)\bigg(-\frac{3 l^8 \Omega_H^2}{z^6}\left(\frac{l^6}{r_+^6}\right) +\frac{l^6 \Omega_H^2 }{z^6} \left(l^2 \left(\Omega_H^2 \left(8 l^2-3 z^2\right)+12\right)+3 z^2\right)\left(\frac{l^4}{r_+^4}\right)\nonumber\\
&&\left.+\cO\left(\frac{l^{2}}{r_+^{2}}\right)
\bigg)\right.\nonumber
\\
{\Omega^{in}_{SG}}_0(z)&=&\frac{l^4 \Omega_H}{z^4}-\frac{l^6 \Omega_H^3}{z^8} \left(l^4-z^4\right)\left(\frac{r_+^2}{l^2}\right) +\frac{l^6 \Omega_H^3}{z^{12}} \left(l^4-z^4\right) \left(l^2 \Omega_H^2 \left(l^4-z^4\right)+z^4\right)\left(\frac{r_+^4}{l^4}\right)\nonumber\\
&&\left. -\frac{l^6 \Omega_H^3}{z^{16}} \left(l^4-z^4\right) \left(l^2 \Omega_H^2 \left(l^4-z^4\right)+z^4\right)^2\left(\frac{r_+^6}{l^6}\right) +\frac{l^6 \Omega_H^3}{z^{20}} \left(l^4-z^4\right) \big(l^2 \Omega_H^2 \right.\nonumber\\
&&\left.\times\left(l^4-z^4\right)+z^4\big)^3\left(\frac{r_+^8}{l^8}\right)+\cO\left(\frac{r_+^{10}}{l^{10}}\right)\right.\nonumber
\\
{\Omega^{in}_{SG}}_2(z)&=&\left(\frac{Q^2}{l^4}\right)\bigg(-\frac{l^4 \Omega_H}{z^6} \left(l^2-z^2\right)\left(\frac{l^4}{r_+^4}\right) +\frac{l^4 \Omega_H}{z^{10}} \left(l^2-z^2\right) \big(l^2 \Omega_H^2 \left(2 l^4+l^2 z^2+z^4\right)\nonumber\\
&&\left.+z^4\big)\left(\frac{l^2}{r_+^2}\right) -\frac{l^4 \Omega_H}{z^{14}} \left(l^2-z^2\right) \Big(z^8 \left(1-l^2 \Omega_H^2 \left(l^2 \Omega_H^2-3\right)\right)+l^4 \Omega_H^2 z^6 \left(3-l^2 \Omega_H^2\right)\right.\nonumber
\\
&&\left.+l^6 \Omega_H^2 \big(l^4 \Omega_H^2 \left(3 l^2+2 z^2\right)
+4 z^4\big)\Big)+\frac{l^4 \Omega_H}{z^{18}} \Big(-z^{14} \big(l^2 \Omega_H^2 \big(l^2 \Omega_H^2 \left(l \Omega_H-2\right) \big(l \Omega_H\right.\nonumber\\
&&\left.+2\big)+6\big)+1\big)+l^2 z^{12}+l^{18} \Omega_H^6 \left(4 l^2-z^2\right)
+3 l^{14} \Omega_H^4 z^4 \left(3-2 l^2 \Omega_H^2\right)-l^{12} \Omega_H^4 z^6 \right.\nonumber\\
&&\left.\times\left(l^2 \Omega_H^2+1\right)+2 l^8 \Omega_H^2 z^8 \left(l^2 \Omega_H^2 \left(l^2 \Omega_H^2-3\right)+3\right)+3 l^8 \Omega_H^4 z^{10} \left(l^2 \Omega_H^2-2\right)\Big)\right.\nonumber\\
&&\left.\times\left(\frac{r_+^2}{l^2}\right) +\frac{l^4 \Omega_H}{z^{22}} \Big(-l^2 z^{16}+l^{24} \Omega_H^8 \left(z^2-5 l^2\right)+4 l^{20} \Omega_H^6 z^4 \left(3 l^2 \Omega_H^2-4\right)
+l^{18}\right.\nonumber\\
&&\left.\times \Omega_H^6 z^6 \left(l^2 \Omega_H^2+1\right)-3 l^{14} \Omega_H^4 z^8 \big(l^2 \Omega_H^2 \left(3 l^2 \Omega_H^2-8\right)+6\big)+l^{12} \Omega_H^4 z^{10} \big(-l^2 \Omega_H^2
\right.\nonumber\\
&&\left.\times
 \left(6 l^2 \Omega_H^2-11\right)
-2\big)+2 l^8 \Omega_H^2 z^{12} \big(l^2 \Omega_H^2 \left(l^2 \Omega_H^2 (l \Omega_H-2) (l \Omega_H+2)+6\right)-4\big)\right.\nonumber\\
&&\left.+l^6 \Omega_H^2 z^{14}\big(l^2 \Omega_H^2 \left(l^2 \Omega_H^2 \left(5 l^2 \Omega_H^2-17\right)+18\right)-2\big)+z^{18} \big(1-l^2 \Omega_H^2 \big(l^6 \Omega_H^6-5 l^4\right.\nonumber\\
&&\left.\times \Omega_H^4+10 l^2 \Omega_H^2-10\big)\big)\Big)\left(\frac{r_+^4}{l^4}\right)+\cO\left(\frac{r_+^{6}}{l^{6}}\right)\bigg)\right.\nonumber
\\
{\Omega^{in}_{SG}}_3(z)&=&\left(\frac{Q^3}{l^6}\right)\bigg(\frac{l^4 \Omega_H}{z^6} \left(l^2-z^2\right)\left(\frac{l^6}{r_+^6}\right) -\frac{l^4 \Omega_H }{z^{10}}\left(l^2-z^2\right) \Big(z^4 \left(3 l^2 \Omega_H^2+2\right)+l^4 \Omega_H^2 \big(3 l^2\nonumber\\
&&\left.+2 z^2\big)\Big)\left(\frac{l^4}{r_+^4}\right) +\frac{l^4 \Omega_H }{z^{14}} \left(l^2-z^2\right) \Big(8 l^4 \Omega_H^2 z^6+z^8 \left(11 l^2 \Omega_H^2+3\right)+l^{10} \Omega_H^4 \big(5 l^2\right.\nonumber\\
&&\left. +4 z^2\big)+l^6 \Omega_H^2 z^4\left(4 l^2 \Omega_H^2+9\right)\Big)\left(\frac{l^2}{r_+^2}\right) -\frac{l^4 \Omega_H}{z^{18}} \Big(-2 z^{14} \left(13 l^2 \Omega_H^2+2\right)+2 l^2 z^{12}\right.\nonumber
\eeqa
\beqa
&&\left.\times \left(3 l^2 \Omega_H^2+2\right)-5 l^{14} \Omega_H^4 z^4 \left(l^2 \Omega_H^2-4\right)+2 l^6 \Omega_H^2 z^{10} \left(2 l^2 \Omega_H^2 \left(l^2 \Omega_H^2-5\right)+1\right)\right.\nonumber\\
&&\left.+l^8 \Omega_H^2 \big(l^6 \Omega_H^4 \left(7 l^6-l^4 z^2-5 z^6\right)+18 z^8\big)\Big)+\frac{l^4 \Omega_H }{z^{22}} \left(l^2-z^2\right) \Big(40 l^4 \Omega_H^2 z^{14}+5 z^{16}\right.\nonumber\\
&&\left. \times\left(10 l^2 \Omega_H^2+1\right)+l^{22} \Omega_H^8 \left(9 l^2+8 z^2\right)
+l^{18} \Omega_H^6 z^4 \left(35-6 l^2 \Omega_H^2\right)-12 l^{16} \Omega_H^6 z^6 \right.\nonumber\\
&&\left.\times\left(l^2 \Omega_H^2-3\right)+l^{12} \Omega_H^4 z^8 \left(50-l^2 \Omega_H^2 \left(7 l^2 \Omega_H^2-11\right)\right)
+4 l^{10} \Omega_H^4 z^{10} \big(l^2 \Omega_H^2 \big(l^2 \Omega_H^2\right.\nonumber\\
&&\left.-6\big)+15\big)+2 l^6 \Omega_H^2 z^{12} \big(2 l^2 \Omega_H^2 \left(l^2 \Omega_H^2 \left(l^2 \Omega_H^2-6\right)+15\right)+15\big)\Big)\left(\frac{r_+^2}{l^2}\right)+\cO\left(\frac{r_+^{4}}{l^{4}}\right)\bigg)\right.\nonumber
\\
{\Omega^{in}_{SG}}_4(z)&=&\left(\frac{Q^4}{l^8}\right)\bigg(
\frac{l^4 \Omega_H}{z^6} \left(z^2-l^2\right)\left(\frac{l^8}{r_+^8}\right) -\frac{l^4 \Omega_H }{z^{12}} \left(l^2-z^2\right) \Big(l^4 \Omega_H^2 \left(l^4-4 l^2 z^2-3 z^4\right)\nonumber\\
&&\left.-3 z^6 \left(2 l^2 \Omega_H^2+1\right)\Big)\left(\frac{l^6}{r_+^6}\right) +\frac{l^4 \Omega_H }{z^{16}} \Big(z^{12} \left(l^2 \Omega_H^2 \left(7 l^2 \Omega_H^2+27\right)+6\right)-2 l^2 z^{10} \right.\nonumber\\
&&\left.\times
\left(l^2 \Omega_H^2 \left(l^2 \Omega_H^2+6\right)+3\right)+l^{10} \Omega_H^2 z^4 \left(l^2 \Omega_H^2+3\right)+3 l^8 \Omega_H^4 \left(l^8-3 l^6 z^2+3 z^8\right)\right.\nonumber\\
&&\left.-9 l^8 \Omega_H^2 z^6 \left(l^2 \Omega_H^2+2\right)\Big)\left(\frac{l^4}{r_+^4}\right)+\cO\left(\frac{l^{2}}{r_+^{2}}\right)\bigg)\right.\nonumber
\eeqa
for the metric functions.  For the gauge field we find
\beqa
{{a_0^{in}}_{SG}}(z)&=&\left(\frac{Q}{l^2}\right)\frac{\sqrt{3} l^2} {z^2}\left(\frac{l}{r_+}\right)\nonumber
\\
{{a_1^{in}}_{SG}}_1(z)&=&\left(\frac{Q}{l^2}\right)\left(-\frac{\sqrt{3} l^4 \Omega_H}{z^2}\left(1 - \left(\frac{r_+^2}{l^2}\right) + \left(\frac{r_+^4}{l^4}\right) - \left(\frac{r_+^6}{l^6}\right) + \left(\frac{r_+^8}{l^8}\right)\right)+\cO\left(\frac{r_+^{10}}{l^{10}}\right)\right)\nonumber
\\
{{a_1^{in}}_{SG}}_2(z)&=&\left(\frac{Q^2}{l^4}\right)\bigg(\frac{\sqrt{3} l^4 \Omega_H}{z^2}\left(\frac{l^2}{r_+^2}\right)-\frac{\sqrt{3} l^4 \Omega_H \left(l^2 \Omega_H^2+2\right)}{z^2}+\frac{3 \sqrt{3} l^4 \Omega_H \left(l^2 \Omega_H^2+1\right)}{z^2}\left(\frac{r_+^2}{l^2}\right)\nonumber\\
&&\left. -\frac{2 \sqrt{3} l^4 \Omega_H \left(3 l^2 \Omega_H^2+2\right)}{z^2}\left(\frac{r_+^4}{l^4}\right)+\cO\left(\frac{r_+^{6}}{l^{6}}\right)\bigg)\right.\nonumber
\\
{{a_1^{in}}_{SG}}_3(z)&=&\left(\frac{Q^3}{l^6}\right)\bigg(-\frac{\sqrt{3} l^4 \Omega_H}{z^2}\left(\frac{l^4}{r_+^4}\right) +\frac{3 \sqrt{3} l^4 \Omega_H \left(l^2 \Omega_H^2+1\right)}{z^2}\left(\frac{l^2}{r_+^2}\right) -\frac{2 \sqrt{3} l^4 \Omega_H}{z^2}\nonumber\\
&&\left. \times\left(l^2 \Omega_H^2 \left(l^2 \Omega_H^2+6\right)+3\right)+\cO\left(\frac{r_+^{2}}{l^{2}}\right)\bigg)\right.\nonumber
\\
{{a_1^{in}}_{SG}}_4(z)&=&\left(\frac{Q^4}{l^8}\right)\bigg(\frac{\sqrt{3} l^4 \Omega_H}{z^2}\left(\frac{l^6}{r_+^6}\right) -\frac{2 \sqrt{3} l^4 \Omega_H \left(3 l^2 \Omega_H^2+2\right)}{z^2}\left(\frac{l^4}{r_+^4}\right)+\cO\left(\frac{l^{2}}{r_+^{2}}\right)\bigg)\nonumber
\eeqa

\end{document}